\documentclass[prd,aps,showpacs,floats,letterpaper,floatfix,superscriptaddress,twocolumn,nofootinbib]{revtex4}

\usepackage{amssymb,amsmath}
\usepackage[dvips]{graphicx}
\usepackage{dcolumn}
\usepackage{hyperref}

\newcommand{\hLN}{\hat{\mathbf{L}}_{\mathrm{N}}}
\newcommand{\hL}{\hat{\mathbf{L}}}
\newcommand{\hS}{\hat{\mathbf{S}}}

\newcommand{\hl}{\hat{\mathbf{\lambda}}}
\newcommand{\hn}{\hat{\mathbf{n}}}
\newcommand{\hN}{\hat{\mathbf{N}}}

\newcommand{\ve}{\mathbf{e}}

\newcommand{\vS}{\mathbf{S}}
\newcommand{\vL}{\mathbf{L}}

\newcommand{\beq}{\begin{equation}}
\newcommand{\eeq}{\end{equation}}
\newcommand{\bea}{\begin{eqnarray}}
\newcommand{\eea}{\end{eqnarray}}
\newcommand{\ba}{\begin{array}}
\newcommand{\ea}{\end{array}}

\newcommand{\braket}[2]{{\langle #1,#2 \rangle}}
\newlength{\sizeonefig}
\newlength{\sizetwofig}
\begin{document}

\title{A physical template family for gravitational waves from
precessing binaries of spinning compact objects: Application to
single-spin binaries}

\author{Yi Pan}

\affiliation{Theoretical Astrophysics and Relativity, California
Institute of Technology, Pasadena, CA 91125}

\author{Alessandra Buonanno}

\affiliation{Groupe de Cosmologie et Gravitation (GReCO), Institut 
d'Astrophysique de Paris (CNRS), 98$^{\rm bis}$ Boulevard Arago, 
75014 Paris, France}

\author{Yanbei Chen}

\affiliation{Theoretical Astrophysics and Relativity, California
Institute of Technology, Pasadena, CA 91125}

\author{Michele Vallisneri}

\affiliation{Jet Propulsion Laboratory,
California Institute of Technology, Pasadena, CA 91109}

\begin{abstract}
  The detection of the gravitational waves (GWs) emitted by precessing
  binaries of spinning compact objects is complicated by the large
  number of parameters (such as the magnitudes and initial directions
  of the spins, and the position and orientation of the binary with
  respect to the detector) that are required to model accurately the
  precession-induced modulations of the GW signal.  In this paper we
  describe a fast matched-filtering search scheme for precessing
  binaries, and we adopt the physical template family proposed by
  Buonanno, Chen, and Vallisneri [\textit{Phys.\ Rev.\ D} \textbf{67},
  104025 (2003)] for ground-based interferometers. This family
  provides essentially exact waveforms, written directly in terms of
  the physical parameters, for binaries with a single significant
  spin, and for which the observed GW signal is emitted during the
  phase of adiabatic inspiral (for LIGO-I and VIRGO, this corresponds
  to a total mass $M \lesssim 15 M_\odot$).  We show how the detection
  statistic can be maximized automatically over all the parameters
  (including the position and orientation of the binary with respect
  to the detector), except four (the two masses, the magnitude of the
  single spin, and the opening angle between the spin and the orbital
  angular momentum), so the template bank used in the search is only
  four-dimensional; this technique is relevant also to the searches
  for GW from extreme--mass-ratio inspirals and supermassive blackhole
  inspirals to be performed using the
  space-borne detector LISA.  Using the LIGO-I design sensitivity, we
  compute the detection threshold ($\sim 10$) required for a
  false-alarm probability of $10^{-3}$/year, and the number of
  templates ($\sim$ 76,000) required for a minimum match of 0.97, for
  the mass range $(m_1,m_2) = [7,12]M_\odot \times [1,3]M_\odot$.
\end{abstract}


\date{\today}

\maketitle

\section{Introduction}
\label{sec:introduction}

Binaries consisting of a black hole (BH) in combination with another
BH or with a neutron star (NS) are among the most promising
gravitational-wave (GW) sources for first-generation
laser-interferometer GW detectors such as LIGO \cite{LIGO,S1}, VIRGO
\cite{VIRGO}, GEO\,600 \cite{GEO,S1} and TAMA\,300 \cite{TAMA}.  For
LIGO-I and VIRGO, and for binaries with total mass $M \lesssim 15
M_\odot$, the observed GW signal is emitted during the
adiabatic-inspiral regime, where post-Newtonian (PN) calculations can
be used to describe the dynamics of the binary and predict the
gravitational waveforms emitted \cite{2PN,DJS,BFIJ,DJSd}.

Very little is known about the statistical distribution of BH spin
magnitudes in binaries: the spins could very well be large, with a
significant impact on both binary dynamics and gravitational waveforms.  On the
contrary, it is generally believed that NS spins will be small in the
NS--BH and NS--NS binaries that are likely to be observed with
first-generation GW detectors. For example, the observed NS--NS binary
pulsars have rather small spin, $S_{\rm NS}/m_{\rm NS}^2 \sim
10^{-3}$~\cite{2PN}. One reason the NSs in binaries of interest for GW
detectors should carry small spin is that they are old enough to have
spun down considerably (even if they once had spins comparable to the
theoretical upper limits, $S_{\rm NS}/m_{\rm NS}^2 \simeq 0.6
\mbox{--}0.7$ \cite{UL}, where $m_\mathrm{NS}$ is the NS mass, and
where we set $G = c = 1$), and because dynamical evolution cannot spin
them up significantly (even during the final phase of inspiral when
tidal torques become important \cite{BC}).

Population-synthesis studies \cite{K00} suggest that in NS--BH
binaries there is a significant possibility for the BH spin to be
substantially misaligned with the orbital angular momentum of the binary.
Early investigations \cite{ACST94,apostolatos2} showed that when
this is the case and the BH spin is large, the evolution of the GW
phase and amplitude during the adiabatic inspiral is significantly
affected by spin-induced modulations. While reliable templates
for precessing binaries should include these modulational effects,
performing GW searches with template families that include all the
\emph{prima facie} relevant parameters (the masses, the spins, the
angles that describe the relative orientations of detector and binary,
and the direction of propagation of GWs to the detector) is
extremely computationally intensive.

Several authors have explored this issue, and they have proposed
\emph{detection template families} (DTFs) that depend on fewer
parameters and that can still reproduce well the expected physical
signals.  An interesting suggestion, built on the results obtained in
Ref.\ \cite{ACST94}, came from Apostolatos \cite{apostolatos2}, who
introduced a modulational sinusoidal term (the {\it Apostolatos
ansatz}) in the frequency-domain phase of the templates to capture the
effects of precession. This suggestion was tested further by
Grandcl\'ement, Kalogera and Vecchio \cite{GKV}.  The resulting
template family has significantly fewer parameters, but its
computational requirements are still very high, and its signal-fitting
performance is not very satisfactory; Grandcl\'ement and Kalogera
\cite{GK} subsequently suggested a modified family of \emph{spiky}
templates that fit the signals better.

After investigating the dynamics of precessing binaries, Buonanno,
Chen and Vallisneri \cite[henceforth BCV2]{bcv2} proposed a new
convention for quadrupolar GW emission in such binaries, whereby the
oscillatory effects of precession are isolated in the evolution of the
GW polarization tensors.  As a result, the response of the detector to
the GWs can be written as the product of a carrier signal and a
modulational correction, which can be handled using an extension of
the Apostolatos ansatz. On the basis of these observations, BCV2 built
a modulated frequency-domain DTF that, for maximal spins, yields
average fitting factors ($\overline{\mathrm{FF}}$, see Sec.\ VIB of
BCV2) of $\simeq 0.97$ for $(7+5)M_\odot$ BH--BH binaries, and $\simeq
0.93$ for $(10+1.4)M_\odot$ NS--BH binaries (see also Tab.\ VIII,
Tab.\ IX, and Fig.\ 14 of BCV2).  Note that the
stationary-phase-approximation (SPA) templates developed for
nonspinning binaries give much lower $\overline{\mathrm{FF}}$s of
$\simeq 0.90$ for $(7+5)M_\odot$ BH--BH binaries, and $\simeq 0.78$
for $(10+1.4)M_\odot$ NS--BH binaries, while according to our
computations the Apostolatos templates give $\overline{\mathrm{FF}}
\simeq 0.81$ for $(10+1.4)M_\odot$ NS--BH binaries.\footnote{
The authors of Refs.\ \cite{GKV,GK} do not include a Thomas precession
term in the physical model used to test the templates; for this
reason, the fitting factors quoted in Refs.\ \cite{GKV,GK} are
substantially lower than our result. Those authors are
currently investigating the effect of that term~ \cite{Gpc}.}

An important feature of the BCV2 templates is that their mathematical
structure allows an automatic search over several of the
modulational parameters (in strict analogy to the automatic search
over initial orbital phase in GW searches for nonspinning binaries),
reducing significantly the number of templates in the search banks,
and therefore the computational cost. However, since many more signal
shapes are effectively (if implicitly) tested against the detector
output, the detection threshold for this DTF should be set higher than
those for simpler families (for the same false-alarm probability).
According to simple false-alarm 
computations performed with Gaussian, stationary
detector noise (see BCV2) for a single template, 
 the gain in FF is larger than the increase
in the threshold only for binaries (such as NS--BH binaries) with low
symmetric mass ratios $m_1 m_2 / (m_1 + m_2)^2$; while the opposite is
true for high mass ratios. [Ultimately, the issue of FF gain versus
threshold increase will be settled only after constructing the
mismatch metric for this template family and performing Monte Carlo
analyses of false-alarm statistics for the entire template bank under
realistic detector noise.]
Although the improvement in FF with the BCV2 DTF is relevant, it is
still not completely satisfactory, because it translates to a loss of
$\sim 20\%$ in detection rate (for the maximal-spin case) with respect
to a perfect template bank (the loss will be higher if the higher
required threshold is taken into account).  Current estimates of
binary-inspiral event rates within the distance accessible to
first-generation GW interferometers hovers around one event per year,
so a reduction of $\sim 20\%$ in the detection rate may not be acceptable.

BCV2 also proposed, but did not test, a new promising family of {\it
physical} templates (i.\ e., templates that are exact within the
approximations made to write the PN equations) for binaries where only
one of the two compact bodies carries a significant spin. This family
has two remarkable advantages: (i) it consists only of the physical
waveforms predicted by the PN equations in the adiabatic limit, so it
does not raise the detection threshold unnecessarily by including
unphysical templates, as the BCV2 DTF did; (ii) all the template
parameters except four are {\it extrinsic}: that is, they can be
searched over semi-algebraically without having to compute all of the
corresponding waveforms.

In this paper we describe a data-analysis scheme that employs this
family, and we estimate the number of templates required for a NS--BH
search with LIGO-I: we assume $1 M_\odot < m_\mathrm{NS} < 3 M_\odot$,
and $7 M_\odot < m_\mathrm{BH} < 12 M_\odot$ (see Sec.\ 
\ref{subsec4:template}).  In a companion paper \cite{pbcv3}, we show
how a simple extension of this template family can be used to search
for the GWs emitted by binaries when both compact bodies have
significant spins (and where of course the adiabatic limit of the PN
equations is still valid). The problem of estimating the parameters of
the binaries is examined in a forthcoming paper \cite{pbcv2}.
  
This paper is organized as follows. In Sec.\ \ref{sec:refresher} we
review the formalism of matched-filtering GW detection, and we
establish some notation.  In Sec.\ \ref{sec:template} we review the PN
dynamics and GW generation in single-spin binaries, and we discuss the
accuracy of the resulting waveforms, indicating the range of masses
to which our physical template family can be applied.  In Sec.\
\ref{sec:parametrization} we describe the parametrization of the
templates, and we discuss the semialgebraic maximization of
signal--template correlations with respect to the extrinsic
parameters.  In Sec.\ \ref{sec:testing} we describe and test a fast
two-stage detection scheme that employs the templates, and we
discuss its false-alarm statistics. In Sec.\ \ref{sec:metric} we build
the template mismatch metric, and we evaluate the number of templates
required for an actual GW search. Finally, in Sec.\ \ref{sec:conclusion}
we summarize our conclusions.

\section{A brief refresher on matched-filtering GW detection}
\label{sec:refresher}

We refer the reader to Ref.\ \cite{bcv1} (henceforth BCV1), for a
self-contained discussion of matched-filtering techniques for GW
detection, which includes all relevant bibliographic references. In
this section we shall be content with introducing cursorily the
quantities and symbols used throughout this paper.

\emph{Matched filtering}
\cite{Wainstein,DA91,Finn,FC,CF94,DS94,S94,ApostolatosFF,O,BSD,FH,DIS,OS99,DIS3} is the
standard method to detect GW signals of known shape, whereby we
compare the detector output with \emph{templates} that approximate
closely the signals expected from a given class of sources, for a
variety of source parameters. The goodness of fit between the template
$h(\lambda^A)$ (where $\lambda^A$ denotes all the source parameters)
and the real GW signal $s$ is quantified by the \emph{overlap}
\begin{equation}
\label{eq:overlap}
\rho[s,h(\lambda^A)] = \frac{\langle s, h(\lambda^A) \rangle}{\sqrt{\langle h(\lambda^A), h(\lambda^A) \rangle}}
\end{equation}
[also known as the \emph{signal-to-noise ratio} after filtering $s$ by
$h(\lambda^A)$], where the inner product $\langle g(t)$, $h(t)
\rangle$ of two real signals with Fourier transforms $\tilde{g}(f)$,
$\tilde{h}(f)$ is given by \cite{CF94}
\begin{equation}
\langle g, h \rangle =
2 \int_{-\infty}^{+\infty} \frac{\tilde{g}^*(f)
\tilde{h}(f)}{S_n(|f|)} df =
4 \, \mathrm{Re} \int_{0}^{+\infty}
\frac{\tilde{g}^*(f) \tilde{h}(f)}{S_n(f)} df;
\end{equation}
throughout this paper we adopt the LIGO-I one-sided noise power
spectral density $S_n$ given by Eq.\ (28) of BCV1. Except where
otherwise noted, we shall always consider normalized templates
$\hat{h}$ (where the hat denotes normalization), for which $\langle
\hat{h}(\lambda^A), \hat{h}(\lambda^A) \rangle = 1$, so we can drop
the denominator of Eq.\ \eqref{eq:overlap}.

A large overlap between a given stretch of detector output and a
particular template implies that there is a high probability that a GW
signal similar to the template is actually present in the output, and
is not being merely simulated by noise alone. Therefore the overlap
can be used as a \emph{detection statistic}: we may claim a detection
if the overlap rises above a \emph{detection threshold} $\rho^*$,
which is set, on the basis of a characterization of the noise, in such
a way that false alarms are sufficiently unlikely.

The maximum (\emph{optimal}) overlap that can be achieved for the signal $s$
is $\sqrt{\langle s, s \rangle}$ (the \emph{optimal signal-to-noise
  ratio}), which is achieved by a perfect (normalized) template $\hat{h} \equiv s /
\sqrt{\langle s, s \rangle}$. In practice, however, this value will
not be reached, for two distinct reasons. First, the template family
$\{\hat{h}(\lambda^A)\}$ might not contain a faithful representation of the
physical signal $w$. The fraction of the theoretical maximum overlap
that is recovered by the template family is quantified by the
\emph{fitting factor} \cite{ApostolatosFF}
\begin{equation}
\mathrm{FF} = \frac{\max_{\lambda^A} \langle w, \hat{h}(\lambda^A) \rangle}{\sqrt{\langle w, w \rangle}}.
\end{equation}
Second, in practice we will usually not be able to use a
\emph{continuous} template family $\{\hat{h}(\lambda^A)\}$, but instead we
will have to settle with a discretized template {bank}
$\{\hat{h}(\lambda_{(k)}^A)\}$, where $(k)$ indexes a finite lattice in
parameter space; so the best template to match a given physical signal
will have to be replaced by a nearby template in the bank. [As we
shall see in Sec.\ \ref{sec:parametrization}, there is a partial
exception to this rule: we can take into account all possible values
of certain parameters, known as \emph{extrinsic parameters}
\cite{S94,O}, without actually laying down templates in the bank along
that parameter direction.]  The fraction of the optimal overlap that
is recovered by the template bank, in the worst possible case, is
quantified by the \emph{minimum match} \cite{DA91,O}. Assuming that the physical
signal belongs to the continuous template family $\{\hat{h}(\lambda^A)\}$,
the minimum match is equal to
\begin{equation}
\mathrm{MM} = \min_{{\lambda'}^A} \max_{(k)} \langle \hat{h}({\lambda'}^A), \hat{h}(\lambda_{(k)}^A) \rangle.
\end{equation}
The required minimum match $\mathrm{MM}$ sets the allowable coarseness
of the template bank \cite{DA91,DS94,S94}: the closer to one the $\mathrm{MM}$, the closer
to one another the templates will need to be laid down. In Sec.\ 
\ref{sec:metric} we shall use a notion of \emph{metric}
\cite{BSD,O,OS99} in parameter space to characterize the
size and the geometry of the template bank corresponding to a given
$\mathrm{MM}$.

\section{Adiabatic Post-Newtonian model for single-spin binary inspirals}
\label{sec:template}

In this section we discuss PN adiabatic dynamics and GW generation for
NS--BH and BH--BH binaries. Specifically, in Secs.\
\ref{subsec1:template}--\ref{subsec3:template} we review the PN
equations and the GW emission formalism developed in BCV2. In Sec.\
\ref{subsec4:template} we extend that analysis to study the accuracy
of the waveforms, and we determine the mass range where the waveforms
produced by adiabatic models can be considered accurate for the
purpose of GW detection. In Sec.\ \ref{subsec5:template} we
investigate the effects of quadrupole--monopole interactions (tidal
torques) on the waveforms. In this paper we restrict our analysis to
binaries where only one body has significant spin, leaving a similar
study of generic binaries to a companion paper \cite{pbcv2}.  As a
further restriction, we consider only binaries in circular orbits,
assuming that they have already been circularized by radiation
reaction as they enter the frequency band of ground-based GW
detectors.

For all binaries, we denote the total mass by $M = m_1 + m_2$ and the
symmetric mass ratio by $\eta = m_1 m_2 / M^2$; we also assume that
the heavier body (with mass $m_1 \ge m_2$) carries the spin
$S_1=\chi_1 m_1^2$, with $0\leq\chi_{1}\leq 1$ (here and throughout
this paper we set $G=c=1$).

\subsection{The PN dynamical evolution}
\label{subsec1:template}

In the adiabatic approach [\onlinecite{KWW},\onlinecite{K},\onlinecite{2PN}] to the evolution of
spinning binaries, one builds a sequence of precessing (due to spin
effects) and shrinking (due to radiation reaction) circular
orbits. The orbital frequency increases as the orbit shrinks.  The
timescales of the precession and shrinkage are both long compared to
the orbital period (this is the \emph{adiabatic condition}), until the
very late stage of binary evolution.  [These orbits are sometimes also
called {\it spherical orbits}, since they reside on a sphere with
slowly shrinking radius.]

The radiation-reaction--induced evolution of frequency can be calculated
by using the energy-balance equation,
\beq
\label{adiabatic}
\dot{\omega} =-\frac{{\cal F}}{d {\cal E}/d \omega}\,,
\eeq
where $\mathcal{E}$ is the orbital-energy function, and
$\mathcal{F}$ the GW energy-flux (or luminosity) function.  
Both have been calculated as functions of the orbital frequency
using PN-expansion techniques, and are determined up
to 3.5PN order \cite{DJS,BFIJ,DJSd}; however,  
spin effects have been calculated only up to 2PN
order \cite{KWW}. The resulting evolution equation for $\omega$,
obtained by inserting the PN expansions
of ${\cal E}$ and ${\cal F}$ into Eq.\ \eqref{adiabatic} and
reexpanding [every $(M\omega)^{4/3}$ counts as 1\,PN] is
\begin{widetext}
\begin{multline} 
 \frac{\dot{\omega}}{\omega^2}=\frac{96}{5}\,\eta\,(M\omega)^{5/3}\,
\left \{1-\frac{743+924\,\eta}{336}\,(M\omega)^{2/3} -\left (\frac{1}{12}\,\left 
[\chi_1(\hL_N\cdot\hS_1)\,\left (113\frac{m_1^2}{M^2}+75\eta\right) \right ]
-4\pi \right )(M\omega) \right. \\  
 \left. + \left (\frac{34\,103}{18\,144}+\frac{13\,661}{2\,016}\,\eta+
\frac{59}{18}\,\eta^2 \right )\,(M\omega)^{4/3} 
-\frac{1}{672}\,(4\,159 +15\,876\,\eta)\,\pi\,(M\omega)^{5/3} \right. \\
 \left. + \Bigg[
\left(\frac{16\,447\,322\,263}{139\,708\,800}-\frac{1\,712}{105}\gamma_E+\frac{16}{3}\pi^2\right)+
\left(-\frac{273\,811\,877}{1\,088\,640}+\frac{451}{48}\pi^2-\frac{88}{3}\hat\theta
\right)\eta \right . \\
\left . +\frac{541}{896}\eta^2-\frac{5\,605}{2\,592}\eta^3
-\frac{856}{105}\log\left[16(M\omega)^{2/3}\right] \Bigg] (M\omega)^2
+ \Bigg (
-\frac{4\,415}{4\,032}+\frac{358\,675}{6\,048}\,\eta+\frac{91\,495}{1\,512}\,\eta^2
\Bigg )\,\pi\,(M\omega)^{7/3} \right \}\,,
\label{omegadot}
\end{multline}
\end{widetext}
where $\gamma_E=0.577\ldots$ is Euler's constant.  We denote by $\hL_N
\propto \mathbf{r} \times \mathbf{v}$ the unit vector along the
orbital angular momentum, where $\mathbf{r}$ and $\mathbf{v}$ are the
two-body center-of-mass radial separation and relative velocity,
respectively. $\hL_N$ is also the unit normal vector to the orbital
plane.  Throughout this paper we shall always use hats to denote unit
vectors. (Note for v3 of this paper on gr-qc: Eq.\ (6) is now revised as per
Ref.\ \cite{errata}; the parameter $\hat{\theta}$ has been determined to be 1039/4620 \cite{thetapar}.)

The quantity $\hat{\theta}$ is an undetermined regularization
parameter that enters the GW flux at 3PN order \cite{BFIJ}.  Another
regularization parameter, $\omega_s$, enters the PN expressions of
$\mathcal{E}$ [Eq.\ \eqref{s1}] and $\mathcal{F}$ at 3PN order, and it
has been determined in the ADM gauge~\cite{DJS,DJSd}, but not yet in the harmonic
gauge. However, Eq.\ \eqref{omegadot} does not depend
on $\omega_s$. As in BCV2, we do not include the
(partial) spin contributions to $\dot{\omega}$ at 2.5PN, 3PN, and 3.5PN
orders, which arise from known 1.5PN and 2PN spin terms of ${\cal E}$
and ${\cal F}$. [To be fully consistent one should know the spin terms 
of ${\cal E}$ and ${\cal F}$ at 2.5PN, 3PN and 3.5PN order.]
In Sec.\ \ref{subsec4:template} we shall briefly
comment on the effect of these terms. We ignore also the
quadrupole--monopole interaction, which we discuss in
Sec.\ \ref{subsec5:template}.

The precession equation for the spin is [\onlinecite{K},\onlinecite{ACST94}] 
\beq
\label{S1dot}
\dot{\vS}_1 =
\frac{\eta}{2M}(M\omega)^{5/3}\,\left(4+3\frac{m_2}{m_1}\right)\,\hL_N\, 
\times \vS_1\,,
\eeq
where we have replaced $r \equiv \mathbf{r}$ and $|\vL_N|$ by their
leading-order Newtonian expressions in $\omega$,
\beq
\label{rLofomega}
r=\left(\frac{M}{\omega^2}\right)^{1/3}, \quad
|\vL_N|=\mu\, r^2\omega=\eta\,M^{5/3}\omega^{-1/3}\,.
\eeq
The precession of the orbital plane (defined by the normal vector
$\hL_N$) can be computed following Eqs.\ (5)--(8) of BCV2, and it reads
\beq
\label{Lhdot}
\dot{\hL}_N=\frac{\omega^2}{2M}\,\left(4+3\frac{m_2}{m_1}\right)\,\vS_1\times\hL_N
\equiv \mathbf{\Omega}_{L} \times \hL_N\,.
\eeq
Equations (\ref{omegadot}), (\ref{S1dot}), and (\ref{Lhdot}) describe the
adiabatic evolution of the three variables $\omega$, $\vS_1$ and
$\hL_N$. From those equations it can be easily deduced that the
magnitude of the spin, $S_1=|\vS_1|$, and the angle between the spin
and the orbital angular momentum, $\kappa_1 \equiv\hL_N \cdot
\widehat{\vS}_1$, are conserved during the evolution.

The integration of Eqs.\ (\ref{omegadot}), (\ref{S1dot}) and
(\ref{Lhdot}) should be stopped at the point where the adiabatic
approximation breaks down. This point is usually reached (e.\ g., for
2PN and 3PN orders) when the orbital energy $\mathcal{E}_{n\rm PN}$
reaches a minimum ${d\mathcal{E}_{n\rm PN}}/{d\omega}=0$ (exceptions
occur at Newtonian, 1\,PN and 2.5\,PN orders, as we shall explain in
more detail in Sec.\ \ref{subsec4:template}). We shall call the
corresponding orbit the Minimum Energy Circular Orbit, or MECO.  Up to
3PN order, and including spin--orbit effects up to 1.5PN order, the
orbital energy $\mathcal{E}(\omega)$ reads
[\onlinecite{2PN},\onlinecite{KWW}]\footnote{Equation (11) of BCV2
suffers from two misprints: the spin--orbit and spin--spin terms should
both be divided by $M^2$.}
\begin{widetext}
\begin{multline}
\label{s1}
\mathcal{E}(\omega) = -\frac{\mu}{2}\,(M\omega)^{2/3}\,\left \{
1 - \frac{(9+\eta)}{12}\,(M\omega)^{2/3} + \frac{8}{3} \left ( 1 + 
\frac{3}{4} \frac{m_2}{m_1} \right ) \frac{\hL_N\cdot \vS_1}{M^2}\,
(M\omega) - \frac{1}{24}(81 - 57 \eta + \eta^2)\,(M\omega)^{4/3} \right. \\
\left. +\left [-\frac{675}{64} + \left (\frac{34445}{576}-\frac{205}{96}\pi^2 +
\frac{10}{3}\,\omega_s\right )\eta
-\frac{155}{96}\eta^2-\frac{35}{5184}\eta^3 \right ]\,(M\omega)^{2}\right \}\,.
\end{multline}
\end{widetext}
Henceforth, we assume $\omega_s=0$, as computed in Ref.\ \cite{DJSd}.

\subsection{The precessing convention}
\label{subsec2:template}

BCV2 introduced a new convention to express the gravitational waveform
generated by binaries of spinning compact objects, as computed in the
quadrupolar approximation; here we review it briefly.  At the
mass-quadrupole leading order, the radiative gravitational field
emitted by the quasicircular binary motion reads
\beq
\label{hij}
h^{ij}=\frac{2\mu}{D}\,\left(\frac{M}{r}\right)\,Q_{c}^{ij}\,,
\eeq
where $D$ is the distance between the source and the Earth, and
$Q_c^{ij}$ is proportional to the second time derivative of the
mass-quadrupole moment of the binary,
\beq
\label{Qc}
Q_c^{ij}=2\left[\lambda^i\,\lambda^j-n^i\,n^j\right]\,,
\eeq
with $n^i$ and $\lambda^i$ the unit vectors along the separation
vector of the binary $\mathbf{r}$, and along the corresponding relative
velocity $\mathbf{v}$. In general, these vectors can be written as 
\bea
\label{n_lambda}
&\hn(t)=\mathbf{e}_1(t)\,\cos\Phi(t)+\mathbf{e}_2(t)\,\sin\Phi(t)\,,\\
&\hl(t)=-\mathbf{e}_1(t)\,\sin\Phi(t)+\mathbf{e}_2(t)\,\cos\Phi(t)\,,
\eea
where $\mathbf{e}_1(t)$, $\mathbf{e}_2(t)$, and $\mathbf{e}_3(t) \equiv \hL_N(t)$ 
are orthonormal vectors, and $\mathbf{e}_{1,2}(t)$ forms a basis for the instantaneous 
orbital plane.

The adiabatic condition for a sequence of quasi-spherical orbits 
states that $\dot{\hn} = \omega \hl$, but in general $\dot{\Phi} \neq \omega$. 
The precessing convention introduced by BCV2 is defined by imposing that 
this condition is satisfied (i.\ e., that $\dot{\Phi}=\omega$), and it requires 
that $\ve_{1,2}(t)$ precess alongside $\hL_N$ as 
\beq
\label{eprecess}
\dot{\mathbf{e}}_i(t)=\mathbf{\Omega}_e(t) \times \mathbf{e}_i(t)\,,\quad i=1,2\,, 
\eeq
with
\beq
\label{omegae}
\mathbf{\Omega}_e(t) 
= \mathbf{\Omega}_{L} -
(\mathbf{\Omega}_{ L}\cdot\hL_N )\hL_N
\eeq
[see Eq.\ (\ref{Lhdot}) for the definition of $\mathbf{\Omega}_{L}$]. 
In this convention, the tensor $Q_c^{ij}$ can be written as
\beq
Q^{ij}_c=-2\Bigl([\mathbf{e}_+]^{ij} \cos 2(\Phi +\Phi_0)+ [\mathbf{e}_\times]^{ij} \sin 2
(\Phi+\Phi_0)\Bigr)\,,
\eeq
with $\Phi_0$ an arbitrary initial phase (see below),
and 
\beq
\mathbf{e}_+=\mathbf{e}_1 \otimes \mathbf{e}_1 - \mathbf{e}_2 \otimes \mathbf{e}_2\,,\quad
\mathbf{e}_\times=\mathbf{e}_1 \otimes \mathbf{e}_2 + \mathbf{e}_2 \otimes \mathbf{e}_1\,.
\label{epet}
\eeq

\subsection{The detector response}
\label{subsec3:template}

The response of a ground-based interferometric detector to the GW signal
of Eq.\ \eqref{hij} is given by
\begin{widetext}
\beq
\label{generich}
h=
\underbrace{
-\frac{2\mu}{D}\frac{M}{r}\,
\Bigl( [\ve_+]^{ij}\,\cos2(\Phi+\Phi_0)+[\ve_\times]^{ij}\,\sin2(\Phi +\Phi_0) \Bigr)
}_\mathrm{factor\;Q:\;wave\;generation}
\underbrace{
\Bigl( [\mathbf{T}_+]_{ij}\,F_+ + [\mathbf{T}_\times]_{ij}\,F_\times \Bigr)
}_\mathrm{factor\;P:\;detector\;projection}
\,;
\eeq
the tensors $[\mathbf{T}_{+,\times}]_{ij}$ are defined by
\beq
\label{Tpxnew}
{\bf T}_+ \equiv \ve^{R}_x \otimes \ve^{R}_x-\ve^{R}_y \otimes
\ve^{R}_y\,,\quad {\bf T}_{\times} \equiv \ve^{R}_x \otimes
\ve^{R}_y+\ve^{R}_y \otimes \ve^{R}_x \,,
\eeq
after we introduce the radiation frame
\bea
\mathbf{e}_x^R &=& -\ve_x^{S} \sin\varphi  + \ve_y^{S} \cos\varphi\,,\\
\mathbf{e}_y^R &=& - \ve_x^{S}\,\cos\Theta\cos\varphi
- \ve_y^{S} \cos\Theta \sin\varphi  +  \ve_z^{S} \sin\Theta \,,\\
\mathbf{e}_z^R &=& + \ve_x^{S} \sin\Theta \cos\varphi
+ \ve_y^{S} \sin\Theta \sin\varphi  +  \ve_z^{S} \cos\Theta = \hN\,,
\eea
where the detector lies in the direction $\hN$ with respect to the
binary [for the definitions of the angles $\Theta$ and $\varphi$ see
Fig.\ 1 of BCV2]. For the antenna patterns $F_+$ and $F_\times$ we
have
\beq
\label{Fpxgeneral}
F_{+,\,\times}=\frac{1}{2}[\bar{\mathbf{e}}_{x}\otimes\bar{\mathbf{e}}_{x}-\bar{\mathbf{e}}_{y}€\otimes \bar{\mathbf{e}}_{y}]^{ij}[\mathbf{T}_{+,\,\times}]_{ij}\,,
\eeq
where $\bar{\mathbf{e}}_{x,\,y}$ are the unit vectors along the orthogonal
interferometer arms. More explicitly \cite{FC},
\bea
\label{Fplus}
F_+ &=& \frac{1}{2}(1+\cos^2 \theta)\,\cos 2\phi\,\cos 2 \psi -
\cos \theta\,\sin 2\phi\,\sin2\psi \,,\\
F_\times &=& \frac{1}{2}(1+\cos^2 \theta)\,\cos 2\phi\,\sin 2 \psi +
\cos \theta\,\sin 2\phi\,\cos2\psi
\label{Fcross}
\eea
\end{widetext}
[for the definitions of the angles $\phi$, $\theta$ and $\psi$, please
see Fig.\ 2 of BCV2].

Mathematically, we see that the factor P of Eq.\ \eqref{generich},
which is independent of time, collects only terms that depend on the
position and orientation of the detector, and that describe the
\emph{reception} of GWs; while factor Q collects only terms that
depend on the dynamical evolution of the binary, and that describe the
\emph{generation} of GWs (at least if the vectors $\mathbf{e}_{1,2,3}$
are defined without reference to the detector, as we will do
soon). Using the language of BCV2, in the precessing convention the
\emph{directional} parameters $\Theta$, $\varphi$, $\phi$,
$\theta$, and $\psi$ are isolated in factor P, while the \emph{basic}
and \emph{local} parameters of the binary are isolated in factor Q.

Physically, we see that factor Q evolves along three different
timescales: (i) the orbital period, which sets the GW \emph{carrier}
frequency $2\dot{\Phi}=2\omega$; (ii) the precession timescale at
which the $\mathbf{e}_{+,\times}$ change their orientation in space,
which modulates the GWs; (iii) the radiation-reaction timescale,
characterized by $\omega/\dot{\omega}$, which drives the evolution of
frequency. In the adiabatic regime, the orbital period is the shortest
of the three: so for convenience we shall define the (leading-order)
{\it instantaneous GW frequency} $f_{\rm GW}$ directly from the
instantaneous orbital frequency $\omega$:
\beq
\label{fGW}
f_{\rm GW} \equiv (2\omega)/(2\pi) =\omega/\pi\,.
\eeq

Thus, what parameters are needed to specify Q completely? Equation
\eqref{omegadot} for $\omega(t)$ can be integrated numerically,
starting from an arbitrary $\omega(0)$,\footnote{When templates are
used in actual GW searches, the initial orbital frequency $\omega(0)$
must be chosen so that most of the signal power (i.\ e., the square of
the optimal signal to noise) is accumulated at GW frequencies higher
than the corresponding $f_\mathrm{GW}(0) = \omega(0)/\pi$,
$$\int_{f_\mathrm{GW}(0)}^{+\infty} \frac{\tilde{h}^*(f)
  \tilde{h}(f)}{S_n(|f|)} df \simeq \int_{0}^{+\infty}
\frac{\tilde{h}^*(f) \tilde{h}(f)}{S_n(|f|)} df.$$
For the range of binary masses considered in this paper, and for the
LIGO-I noise curve, such a $f_\mathrm{GW}(0)$ should be about 40
Hz. Most of the calculations performed in this paper (for instance,
the convergence tests and the calculation of the mismatch metric) set
instead $f_\mathrm{GW}(0) = 60$ Hz to save on computational time;
experience has proved that the results are quite stable with respect
to this change.} after we specify the basic parameters $M$, $\eta$,
and $\chi_1$, and the local parameter $\kappa_1 \equiv \hL_N \cdot
\hat{\vS}_1$, which is conserved through evolution.  With the
resulting $\omega(t)$ we can integrate Eqs.\ \eqref{S1dot} and
\eqref{Lhdot}, and then Eq.\ \eqref{eprecess}. For this we need
initial conditions for $\hat{\vS}_1$, $\hLN$, and for the
$\mathbf{e}_i$: without loss of generality, we can introduce a (fixed)
source frame attached to the binary,
\begin{equation}
\begin{gathered}
\label{sourcedef}
\mathbf{e}_{x}^{S}\propto \vS_1(0)-[\vS_1(0)\cdot \hLN(0)]\,\hLN(0)\,, \\
\mathbf{e}_{y}^{S}=\hLN(0)\times\mathbf{e}_{x}^{S}
\,,\quad
\mathbf{e}_{z}^{S}=\hLN(0)\,,
\end{gathered}
\end{equation}
and then take
\beq
\label{eijdef}
\mathbf{e}_{1}(0)=\mathbf{e}_{x}^{S}\,,\quad
\mathbf{e}_{2}(0)=\mathbf{e}_{y}^{S}\,,\quad
\mathbf{e}_{3}(0)=\mathbf{e}_{z}^{S}\,.
\eeq
[If $\vS_1(0)$ and $\hLN(0)$ are parallel, $\mathbf{e}_x^S$ can be
chosen to lie in any direction  orthogonal to $\hLN(0)$.]  
The initial orbital phase $\Phi_0$ that enters the expression of Q is
defined by
\beq
\label{phiinit}
\hn(0) = \mathbf{e}_1(0)\cos\Phi_0 + \mathbf{e}_2(0) \sin\Phi_0\,,
\eeq
while the initial conditions for $\hat{\vS}_1$ and $\hLN$, as expressed by their
components with respect to the source frame, are
\begin{gather}
\label{inithLN}
\hLN(0)=(0,0,1)\,,\\
\label{initS1}
\hat{\vS}_1(0)=\Bigl(\sqrt{1-\kappa_1^2},0,\kappa_1\Bigr)\,.
\end{gather}
BCV2 proposed to use the family of waveforms (detector responses)
defined by Eqs.\
\eqref{omegadot}, \eqref{S1dot}, \eqref{Lhdot}, \eqref{eprecess}, and
\eqref{generich} as a family of \emph{physical templates}
for compact binaries with a single spin. Depending on the maximum PN
order $N$ up to which the terms of Eq.\ (\ref{omegadot}) are retained,
we shall denote this class of template families $\mathrm{ST}_N$. The
$\mathrm{ST}_N$ templates deserve to be called {\it physical} because
they are derived from a physical model, namely the adiabatic PN
dynamics plus quadrupole GW emission.  Each $\mathrm{ST}_N$ template
family is indexed by {\it eleven} parameters: $M$, $\eta$, $\chi_1$
(basic), $\kappa_1$ (local), $\Theta$, $\varphi$, $\theta$, $\phi$,
$\psi$ (directional), plus the initial frequency $\omega(0)$ (or
equivalently, the time $t_0$ at an arbitrary GW frequency), and the
initial phase $\Phi_0$. Of these, using the distinction between
intrinsic and extrinsic parameters introduced in Ref.\
\cite{O}\footnote{Note that the concept of extrinsic and intrinsic
parameters had been present in the data-analysis literature for a long
time (see, e.\ g., \cite{Wainstein}). Sathyaprakash
\cite{S94} draws the same distinction between \emph{kinematical}
and \emph{dynamical} parameters.} and further discussed in BCV2, the
first four are intrinsic parameters: that is, when we search for GWs using
$\mathrm{ST}_N$ templates, we need to lay down a discrete template
bank along the relevant ranges of the intrinsic dimensions. The other
seven are extrinsic parameters: that is, their optimal values can be found
semialgebraically without generating multiple templates along the
extrinsic dimensions (another way of saying this is that the
maximization of the overlap over the extrinsic parameters can be
incorporated in the detection statistic, which then becomes a function
only of the intrinsic parameters). In Sec.\
\ref{sec:parametrization} we shall describe how this maximization over
the extrinsic parameters can be achieved in practice.

\subsection{Comparison between different Post Newtonian orders and
  the choice of mass range}
\label{subsec4:template}
\begin{figure}
\begin{center}
\includegraphics{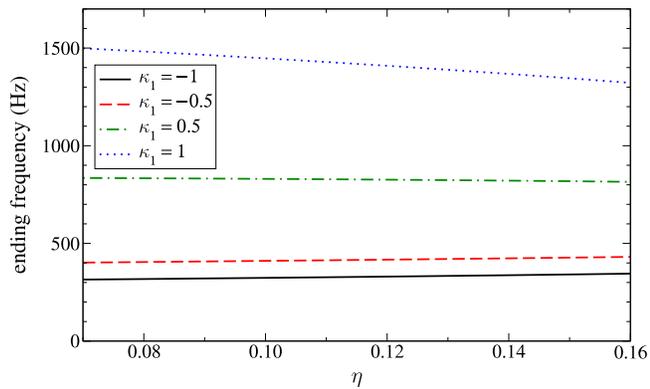} 
\caption{Ending frequency (instantaneous
GW frequency at the MECO) as a function of $\eta$, evaluated from
Eq.\ (\ref{s1}) at 2PN order for $M=15 M_\odot$, $\chi_1 = 1$, and for different
values of $\kappa_1$.\label{Fig0}}
\end{center}
\end{figure}
\begin{figure*}
\begin{center}
\includegraphics{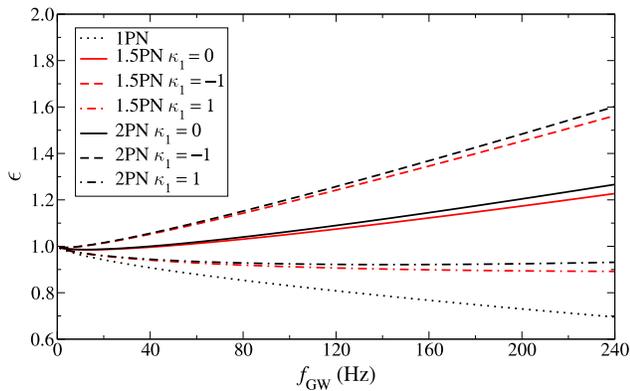} \hspace{0.5cm}
\includegraphics{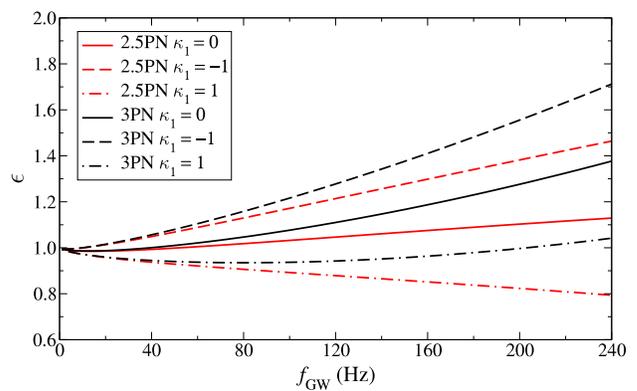} 
\caption{Plot of 
$\epsilon \equiv (\dot{\omega}/\omega^2)/(96/5 \eta
(M\,\omega)^{5/3})$ as a function of $f_{\rm GW} = \omega/\pi$,
evaluated from Eq.\ \eqref{omegadot} at different PN
orders for a $(10 + 1.4)\,M_\odot$ binary.
We do not show the 3.5PN curves, which are very close to the 3PN curves.
\label{Fig1}}
\end{center}
\end{figure*}
\begin{figure}
\begin{center}
\begin{tabular}{c}
\includegraphics{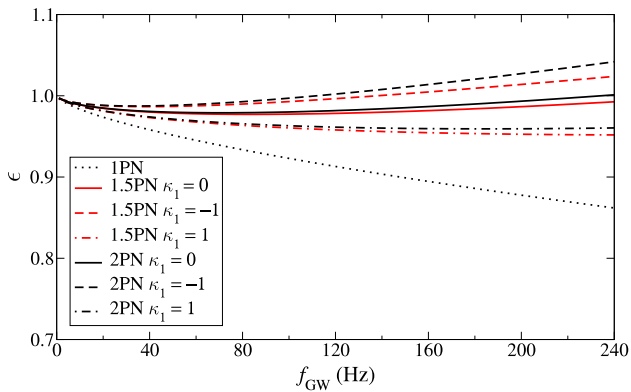} 
\end{tabular}
\caption{Plot of 
$\epsilon \equiv (\dot{\omega}/\omega^2)/(96/5 \eta
(M\,\omega)^{5/3})$ as a function of $f_{\rm GW} = \omega/\pi$,
evaluated from Eq.\ \eqref{omegadot} at different PN
orders for a $(1.4 + 1.4)\,M_\odot$ NS--NS binary.
We do not show the 2.5PN, 3PN ($\widehat{\theta}=0$), and 3.5PN curves,
which are very close to the 2PN curves. Note the change in scale with respect to Fig.\ \ref{Fig1}.
\label{Fig3}}
\end{center}
\end{figure}

In this section we investigate the range of masses $m_1$ and $m_2$ for
which the PN-expanded evolution equations (\ref{omegadot}),
(\ref{S1dot}), and (\ref{Lhdot}) [and therefore the template family
(\ref{generich})] can be considered reliable.  As a rule of thumb, 
we fix the largest acceptable value of the total mass by requiring that the
\emph{GW ending frequency} (in our case, the instantaneous GW
frequency at the MECO) should not lie in the frequency band of good
detector sensitivity for LIGO-I.  Considering the results obtained by
comparing various nonspinning PN models [\onlinecite{DIS3},BCV1], and
considering the variation of the ending frequency when spin effects
are taken into account [BCV2], we require $M \leq 15 M_\odot$. In
keeping with the focus of this paper on binaries with a single
significant spin, we also impose $m_2/m_1 \leq 0.5$, which constrains
the spin of the less massive body to be relatively small (of course, this
condition is always satisfied for NS--BH binaries).  As a matter of
fact, population-synthesis calculations \cite{PS} suggest that the
more massive of the two compact bodies will have the larger spin,
since usually it will have been formed first, and it will have been
spun up through accretion from the progenitor of its companion.  For
definiteness, we assume $m_1 = 1 \mbox{--} 3M_\odot$ and $m_2 = 7
\mbox{--} 12 M_\odot$; the corresponding range of $\eta$ is
$0.07\mbox{--}0.16$.

In Fig.\ \ref{Fig0} we plot the GW ending frequency as a function of
$\eta$, evaluated from Eq.\ (\ref{s1}) at 2PN order for $M=15 M_\odot$
and $\chi_1 = 1$.  The various curves refer to different values of
$\kappa_1$. The minimum of the GW ending frequency is $\sim 300$ Hz,
and it corresponds to a $(12+1)M_\odot$ binary with spin antialigned
with the orbital angular momentum.  In Fig.\ \ref{Fig1} we plot
$\dot{\omega}/\omega^2$, normalized to its leading (Newtonian) term
$96/5 \eta (M\,\omega)^{5/3}$, as a function of the instantaneous GW
frequency; $\dot{\omega}/\omega^2$ is evaluated from Eq.\ 
(\ref{omegadot}) at different PN orders, for a $(10 + 1.4)\,M_\odot$
binary with $\chi_1 = 1$.  We see that the effects of the spin--orbit
interaction (evident for different $\kappa_1$ within the same PN
order) are comparable to, or even larger than, the effect of
increasing the PN order.  We see also that the different PN curves
spread out more and more as we increase $M$ and $\eta$.
For comparison, in Fig.\ \ref{Fig3} we show
the same plot for a $(1.4+1.4)M_{\odot}$ NS--NS binary; note the
different scale on the vertical axis. In this case the various curves
remain rather close over the entire frequency band.

Another procedure (often used in the literature) to characterize the effects
of spin and PN order on the evolution of the GW frequency is to count
the number of GW cycles accumulated within a certain frequency band:
\beq
{\cal N}_{\rm GW} \equiv
\frac{1}{\pi}\int_{\omega_{\rm min}}^{\omega_{\rm max}}\,
\frac{\omega}{\dot{\omega}}\,d\omega\,.
\eeq
Here we take $\omega_{\rm min}=\pi\times 10\,$Hz and $\omega_{\rm
  max}=\omega_{\rm ISCO} = (6^{3/2}\pi M)^{-1}$, corresponding to the
orbital frequency at the innermost stable circular orbit (ISCO) of a
Schwarzchild black hole with mass $M$.  In Table \ref{Tab1} we show
${\cal N}_{\rm GW}$ at increasing PN orders, for $(10+1.4)M_\odot$,
$(12+3)M_\odot$, and $(7+3)M_\odot$ binaries.
The contributions in parentheses are partial spin
terms present at 2.5PN, 3PN, and 3.5PN orders, and due to 
known 1.5PN spin terms in the orbital energy and luminosity. These terms were
neglected in Eq.\ (\ref{omegadot}) to be consistent in including PN
terms of the same order, and we list them here only to give their
order of magnitude. Unless there are cancellations, the large number
of cycles suggests that it is worth to compute spin effects up to the
3.5PN order.
\begin{table*}[t]
\begin{center}
{\scriptsize 
\begin{tabular}{r||D{,}{.}{1}@{\,\,}l@{}D{,}{.}{1}@{}l|D{,}{.}{1}@{\,\,}l@{\,\,}D{,}{.}{1}@{}l|D{,}{.}{1}@{\,\,}l@{\,\,}D{,}{.}{1}@{}l}
& \multicolumn{4}{c|}{$(10+1.4)M_\odot$}
& \multicolumn{4}{c|}{$(12+3)M_\odot$}
& \multicolumn{4}{c}{$(7+3)M_\odot$} \\
\hline\hline
Newtonian       & 3577,0 & &&
                        & 1522,3 & &&
                        &  2283,8 & && \\
1PN                     &  213,1 & &&
                        &   114,3 & &&
                        &   139,0 & && \\
1.5PN           & -181,3 & +& 114,2  & $ \kappa_1\chi_1$
                &  -99,7 & +& 55,7  & $ \kappa_1\chi_1$
                &  -102,3 & +& 48,2  & $ \kappa_1\chi_1$ \\
2PN                     & 9,8  & &&
                        & 6,3  & &&
                        & 6,4  & && \\
2.5PN           & -20,4& + (& 21,1 & ${\kappa_1\chi_1}$)
                        & -12,7 & + (& 12,1  & ${\kappa_1\chi_1}$)
                        & -10,9 & + (& 9,0  & ${\kappa_1\chi_1}$) \\
3PN                     & 2,2  & + (& -17,0  &
${\kappa_1\chi_1}$ + $2.4 \kappa_1^2\chi_1^2$) +0.42\,$\hat\theta$
                        & 2,2  & + (& -9,7  &
                        ${\kappa_1\chi_1}$ + 1.2$\kappa_1^2\chi_1^2$)
+0.40\,$\hat\theta$ 
                        & 2,3  & + (& -6,6  &
                        ${\kappa_1\chi_1}$ + 0.7$\kappa_1^2\chi_1^2$) 
+0.43\,$\hat\theta$
\\
3.5PN           & -1,9 & + (& 6,4  & ${\kappa_1\chi_1}$)
                        & -1,3 & + (& 3,8  & ${\kappa_1\chi_1}$) 
                        & -1,1 & + (& 2,6  & ${\kappa_1\chi_1}$) 
\\
\end{tabular}}\end{center}
\caption{PN contributions to the number ${\cal N}_{\rm GW}$ of GW
cycles accumulated from
$\omega_{\rm min} = \pi\times 10\,{\rm Hz}$ to  
$\omega_{\rm max} = \omega_{\rm ISCO}=1/(6^{3/2}\,\pi\,M)$. 
The contributions in parentheses, ``(...)'', are partial spin
terms present at 2.5PN, 3PN and 3.5PN orders and due to 
known 1.5PN spin terms in the orbital energy and luminosity. 
\label{Tab1}}
\end{table*}

The number of accumulated GW cycles $\mathcal{N}_{\rm GW}$ can be a
useful diagnostic, but taken alone it provides incomplete and
sometimes even misleading information. There are three reasons for
this.  First, $\mathcal{N}_{\rm GW}$ is related only to the number of
orbital cycles of the binary {\it within} the orbital plane, but it
does not reflect the precession of the plane, which modulates the
detector response in both amplitude and phase. These modulations are
very important effects, as witnessed by the fact that neither the
standard nonspinning-binary templates (which do not have builtin
modulations) nor the original Apostolatos templates (which add
only modulations to the phase) can reproduce satisfactorily the
detector response to the GWs emitted by precessing binaries.  Second,
even if two signals have $\mathcal{N}_{\rm GW}$ that differ by $\sim
1$ when $\omega_{\rm max}$ equals the GW ending frequency (which
apparently represents a total loss of coherence, and hence a
significant decrease in overlap), one can always shift their arrival
times to obtain higher overlaps.  Third, in the context of GW
searches the differences in $\mathcal{N}_{\rm GW}$ should be
minimized with respect to the search parameters, {\it \`a la} fitting
factor.

The Cauchy criterion \cite{DIS} implies that the sequence ${\rm
  ST}_N$ converges if and only if, for every $k$, $\braket{{\rm
    ST}_{N+k}}{{\rm ST}_{N}} \rightarrow 1$ as $N \rightarrow \infty$.
One requirement of this criterion is that $\braket{{\rm
    ST}_{N+0.5}}{{\rm ST}_{N}} \rightarrow 1$ as $N \rightarrow
\infty$, and this is what we test in Table \ref{CauchyST123}, for
maximally spinning and nonspinning $(10 + 1.4)M_\odot$ and $(12 +
3)M_\odot$ binaries. The overlaps quoted at the beginning of each
column are maximized over the extrinsic parameters $t_0$ and $\Phi_0$,
but not over the five extrinsic directional parameters $\varphi$,
$\Theta$, $\theta$, $\phi$ and $\psi$ or the intrinsic parameters
$m_1$, $m_2$, $\chi_1$ and $\kappa_1$.  By contrast, we show in
parentheses the overlaps maximized over all the parameters of the
lower-order family (i.\ e., the fitting factors FF for the target
family $\mathrm{ST}_{N+k}$ as matched by the search family
$\mathrm{ST}_N$); we show in brackets the parameters at which the
maximum overlaps are achieved. [The overlaps are especially bad when
1PN and 2.5PN waveforms are used.  These two orders are rather
particular: the flux function $\mathcal{F}$ can be a decreasing
function of $\omega$, and even assume negative values (which is
obviously not physical); correspondingly, $\dot{\omega}$ can
become negative.  Furthermore, the MECO criterion used to set the
ending frequency can also fail, because for some configurations the
MECO does not exist, or occurs after $\dot{\omega}$ has become
negative.  To avoid these problems, we stop the numerical integration
of the equations of motion when $\dot{\omega}$ decreases to one tenth
of its Newtonian value, or at a GW frequency of 1\,kHz, whichever
comes first. For comparison, in Table \ref{CauchyST123} we show also
the overlaps between $\mathrm{ST}_2$ and $\mathrm{ST}_3$, which are
much higher than those between $\mathrm{ST}_2$ and $\mathrm{ST}_{2.5}$,
and than those between $\mathrm{ST}_{2.5}$ and $\mathrm{ST}_3$.]

While the nonmaximized overlaps can be very low, the FFs
are consistently high (note that this requires extending the search
into the unphysical template region where $\eta > 0.25$ and $\chi_1
>1$); however, the best-fit search parameters can be rather different
from the target parameters. This suggests that higher-order PN effects
can be reabsorbed by a change of parameters, so the
$\mathrm{ST}_N$ templates can be considered rather reliable for the
purpose of detecting GWs from precessing binaries in the mass range
examined; however, the estimation of binary parameters can suffer from
systematic errors. In the rest of this paper we shall describe and
analyze a search scheme that uses the $\mathrm{ST}_2$ template family.

A more thorough analysis of the differences between the various PN
orders would be obtained by comparing the PN-expanded adiabatic model
used in this paper with PN-resummed adiabatic models ({\it \`a la} Pad\'e
\cite{DIS}) and nonadiabatic models ({\it \`a la} effective-one-body
\cite{EOB}). A similar comparison was carried out for the nonspinning
case in Refs.\ [\onlinecite{DIS3},BCV1]. Unfortunately, waveforms that
include precessional effects are not yet available for the PN-resummed
adiabatic and nonadiabatic models.
\begin {table*} 
\begin {tabular}{l||rr|rr|rr|rr}
\multicolumn{1}{c||}{($N+k$,$N$)} & \multicolumn{8}{c}{$\left <{\rm
ST}_{N+k},{\rm ST}_{N} \right >$ for $(10+1.4)M_\odot$ binary, $\chi_1 = 1$} \\
     & \multicolumn{2}{c|}{$\kappa_1 = 0.9$}
     & \multicolumn{2}{c|}{$\kappa_1 = 0.5$}
     & \multicolumn{2}{c|}{$\kappa_1 = -0.5$}
     & \multicolumn{2}{c}{$\kappa_1 = -0.9$}\\
\hline\hline
 (1,0)     & 0.1976 & (0.7392) [24.5,0.02,0.00,0.00]  & 0.1976 & (0.7392)  & 0.1976 & (0.7392) [24.5,0.02,0.00,$-$0.00] & 0.1976 & (0.7392)  \\
 (1.5,1)   & 0.2686 & (0.7848) [4.53,0.54,0.00,0.00]  & 0.2696 & (0.7008)  & 0.2065 & (0.6040) [6.58,0.36,0.00,$-$0.00] & 0.1800 & (0.6255) \\
 (2,1.5)   & 0.4876 & (0.99) [9.56,0.14,0.83,0.93]  & 0.5627 & ( 0.99)  & 0.6623 & (0.99) [11.7,0.10,0.97,$-$0.50] & 0.7728 & (0.9760) \\
 (2.5,2)   & 0.1587 & (0.9578) [10.5,0.13,1.56,0.95] & 0.2011 & (0.9887)  & 0.2902 & (0.9398) [10.2,0.13,2.00,$-$0.19] & 0.3460 & (0.99) \\
 (3,2)     & 0.4395 & (0.9848) [11.5,0.10,0.84,0.81] & 0.5057 & (0.9881)  & 0.5575 & (0.9712) [12.0,0.10,0.92,$-$0.48] & 0.6606 & (0.99)\\
 (3,2.5)   & 0.1268 & (0.9758) [12.8,0.08,0.05,0.98] & 0.1539 & (0.99)  & 0.2520 & (0.9744) [25.6,0.03,0.35,$-$0.21] & 0.2488 & (0.99)\\
 (3.5,3)   & 0.9614 & (0.99) [11.7,0.10,1.00,0.90] & 0.9738 & (0.99)  & 0.9907 & (0.99) [11.3,0.11,1.02,$-$0.49] & 0.9939 & (0.99)
\end {tabular}

\vspace{0.5cm} 

\begin {tabular}{l||rr|rr|rr|rr}
\multicolumn{1}{c||}{($N+k$,$N$)} & \multicolumn{8}{c}{$\left <{\rm
ST}_{N+k},{\rm ST}_{N} \right >$ for $(12+3)M_\odot$ binary, $\chi_1 = 1$} \\
     & \multicolumn{2}{c|}{$\kappa_1 = 0.9$}
     & \multicolumn{2}{c|}{$\kappa_1 = 0.5$}
     & \multicolumn{2}{c|}{$\kappa_1 = -0.5$}
     & \multicolumn{2}{c}{$\kappa_1 = -0.9$}\\
\hline\hline
 (1,0)     & 0.2506 & (0.7066) [10.5,0.22,0.00,0.00] & 0.2506 & (0.7066)  & 0.2506 & (0.7066) [10.5,0.22,0.00,$-$0.00] & 0.2506 & (0.7066)  \\
 (1.5,1)   & 0.3002 & (0.7788) [8.22,0.50,0.00,0.00] & 0.2597 & (0.7381)  & 0.2124 & (0.6934) [11.6,0.44,0.00,$-$0.00] & 0.2017 & (0.5427) \\
 (2,1.5)   & 0.6379 & (0.99) [16.0,0.14,1.14,0.92] & 0.7089 & (0.99)  & 0.8528 & (0.99) [14.2,0.18,1.14,$-$0.59] & 0.8620 & (0.99) \\
 (2.5,2)   & 0.2039 & (0.9397) [15.4,0.17,1.95,0.87] & 0.2800 & (0.9863)  & 0.4696 & (0.9756) [13.5,0.18,1.22,$-$0.51] & 0.4219 & (0.99) \\
 (3,2)     & 0.6679 & (0.9851) [11.0,0.25,0.72,0.84] & 0.7267 & (0.99)  & 0.9052 & (0.99) [13.1,0.21,1.50,$-$0.70] & 0.8868 & (0.99)\\
 (3,2.5)   & 0.1603 & (0.99) [18.5,0.10,0.05,0.99] & 0.2272 & (0.99)  & 0.3804 & (0.9759) [15.8,0.15,0.94,$-$0.49] & 0.3060 & (0.99)\\
 (3.5,3)   & 0.9517 & (0.99) [15.2,0.15,0.84,0.86] & 0.9694 & (0.99)  & 0.9932 & (0.99) [15.3,0.16,1.00,$-$0.49] & 0.9900 & (0.99)
\end {tabular}

\vspace{0.5cm}

\begin {tabular}{l||rr|rr}
\multicolumn{1}{c||}{($N+k$,$N$)} & \multicolumn{4}{c}{$\left <{\rm ST}_{N+k},{\rm ST}_{N} \right >$ for $\chi_1=0$}\\ 
     &  \multicolumn{2}{c|}{$(10+1.4)M_\odot$} 
     &  \multicolumn{2}{c}{$(12+3)M_\odot$} \\
\hline\hline
 (1,0)     & 0.1976 & (0.7392) [24.5,0.02] & 0.2506 & (0.7066) [10.5,0.22] \\
 (1.5,1)   & 0.1721 & (0.6427) [5.22,0.51] & 0.2153 & (0.6749) [9.22,0.51] \\
 (2,1.5)   & 0.7954 & (0.9991) [12.7,0.09] & 0.8924 & (0.9981) [16.2,0.14] \\
 (2.5,2)   & 0.4872 & (0.9961) [6.94,0.25] & 0.5921 & (0.9977) [8.05,0.48] \\
 (3,2)     & 0.7471 & (0.9970) [15.3,0.06] & 0.8982 & (0.9994) [19.3,0.10] \\
 (3,2.5)   & 0.4127 & (0.9826) [26.5,0.02] & 0.5282 & (0.9783) [29.0,0.05] \\
 (3.5,3)   & 0.9931 & (0.99) [11.6,0.11] & 0.9924 & (0.99) [15.4,0.15]
\end {tabular}

\caption{\label{CauchyST123}
  Test of Cauchy convergence of the adiabatic templates
  $\mathrm{ST}_N$ at increasing PN orders, for $(10+1.4)M_{\odot}$ and
  $(12+3)M_{\odot}$ binaries, and for maximally spinning BHs
  ($\chi_1=1$, upper and middle panels) and nonspinning BHs
  ($\chi_1=0$, lower panel). The overlaps quoted at the beginning of
  each column are maximized only with respect to the extrinsic
  parameters $t_0$ and $\Phi_0$. In parentheses, ``(...)'', we give
  the overlaps maximized over all the parameters of the lower-order
  family (i.\ e., the fitting factors FF for the target family
  $\mathrm{ST}_{N+k}$ as matched by the search family $\mathrm{ST}_N$,
  evaluated at the target masses shown); the fitting factors are
  obtained by extending the search into the unphysical template region
  where $\eta > 0.25$ and $\chi_1 >1$. In brackets, ``[...]'', we show
  the parameters $M,\eta,\chi_1,\kappa_1$ (or $M,\eta$ when $\chi_1 =
  0$) at which the maximum overlap is achieved.  The detector is
  perpendicular to the initial orbital plane, and at 3PN order we set
  $\hat\theta=0$; in all cases the integration of the equations is
  started at a GW frequency of $60$ Hz.}
\end {table*}

\subsection{The quadrupole--monopole interaction}
\label{subsec5:template}

In this section we investigate the effect of the quadrupole--monopole
interaction, which we have so far neglected in describing the dynamics
of precessing binaries.  It is well known \cite{G} that the quadrupole
moment of a compact body in a binary creates a distortion in its
gravitational field, which affects orbital motion (both in the
evolution of $\omega$ and in the precession of $\hL_N$), and therefore
GW emission; the orbital motion, on the other hand, exerts a torque on
the compact body, changing its angular momentum (i.\ e., it induces a
{\it torqued precession}).  Although the lowest-order
quadrupole--monopole effect is Newtonian, it is smaller than
spin--orbit effects and of the same order as spin--spin effects.

When the the spinning body is a black hole, the equations for the
orbital evolution and GW templates are modified as follows to include
quadrupole--monopole effects.
Eq.\ (\ref{omegadot}) gets the additional term \cite{QM}
\begin{widetext}
\beq
\left (\frac{\dot{\omega}}{\omega^2} \right )_{\rm Quad-Mon} 
=\frac{96}{5}\,\eta\,(M\omega)^{5/3}\,\left \{\frac{5}{2}\chi_1^2\,\frac{m_1^2}{M^2}\,
\left [ 3 (\hL_N \cdot \hS_1)^2-1 \right ]\,(M\omega)^{4/3}
\right \}\,,
\eeq 
while the precession equations (\ref{S1dot})--(\ref{Lhdot}) become \cite{QM}  
\beq
\dot{\vS}_1 = \frac{\eta}{2M}\,(M\omega)^{5/3}\,\left [
\left(4+3\frac{m_2}{m_1}\right) 
- 3\chi_1 (M\omega)^{1/3}\, (\hL_N \cdot \hS_1)\right ]\, 
(\hL_N\,\times \vS_1)\,, 
\eeq
and 
\beq
\label{LhdotQM}
\dot{\hL}_N  
=\frac{\omega^2}{2M}\,\left [ 
\left(4+3\frac{m_2}{m_1}\right) - 3\chi_1(M\,\omega)^{1/3}\,(\hL_N \cdot \hS_1)
\right ]\,(\vS_1\times\hL_N)
\equiv \mathbf{\Omega}'_{L} \times \hL_N \,;
\eeq
furthermore, the orbital energy (\ref{s1}) gets the additional term
\beq
E_{\rm Quad-Mon}(\omega) = -\frac{\mu}{2}\,(M\omega)^{2/3}\,
\left \{-\frac{1}{2}\chi_1^2\,\frac{m_1^2}{M^2}
\left [ 3 (\hL_N \cdot \hS_1)^2-1 \right ]\,(M\omega)^{4/3}\right \}\,;
\eeq
\end{widetext}
last, $\mathbf{\Omega}_e$ is again obtained from Eq.\ (\ref{omegae}),
using the modified $\mathbf{\Omega}'_{L}$ in Eq.\ (\ref{LhdotQM}).

The quadrupole--monopole interaction changes the number of GW cycles
listed in Table \ref{Tab1} at 2PN order. The additional contributions are
$5.2\,\chi_1^2 -15.5\,\kappa_1^2\,\chi_1^2$ for a $(10+1.4) M_\odot$
binary, $2.5\,\chi_1^2 -7.6\,\kappa_1^2\,\chi_1^2$ for a $(12+3)
M_\odot$ binary, and $1.8\,\chi_1^2 -5.4\,\kappa_1^2\,\chi_1^2$ for a
$(7+3) M_\odot$ binary.  To estimate more quantitatively the effect of
the quadrupole--monopole terms, we evaluate the unmaximized overlaps
between 2PN templates, computed with and without the new terms.  The
results for $(10+1.4) M_\odot$ binaries are summarized in Table
\ref{tab:QM}. In parentheses we show the fitting factors, which are
all very high; in brackets we show the intrinsic parameters at which
the maximum overlaps are obtained.  We conclude that for the purpose
of GW searches, we can indeed neglect the effects of the
quadrupole--monopole interaction on the dynamical evolution of the
binary.
\begin{table*} 
\begin{center}
\begin{tabular}{l||rr|rr|rr|rr}
\multicolumn{1}{c||}{{\rm view}} & \multicolumn{8}{c}{$(10+1.4)M_\odot$ with $\chi_1 = 1$}\\
     & \multicolumn{2}{c|}{$\kappa_1 = 0.9$}
     & \multicolumn{2}{c|}{$\kappa_1 = 0.5$}
     & \multicolumn{2}{c|}{$\kappa_1 = -0.5$}
     & \multicolumn{2}{c}{$\kappa_1 = -0.9$}\\
\hline\hline
 {\rm top}      & 0.4796 & (0.99) [10.3,0.13,1.21,0.89] & 0.9890 & (0.99)  & 0.1873 & (0.99) [11.3,0.11,1.08,$-$0.48] & 0.7245 & (0.9877)  \\
 {\rm side}     & 0.3503 & (0.99) [10.0,0.13,0.77,0.94] & 0.8033 & (0.99)  & 0.8754 & (0.99) [11.4,0.11,1.03,$-$0.39] & 0.7598 & (0.99) \\
 {\rm diagonal} & 0.3292 & (0.99) [11.2,0.11,0.80,0.94] & 0.6669 & (0.99)  & 0.4546 & (0.99) [11.3,0.11,1.08,$-$0.49] & 0.8437 & (0.9887) \\
\end{tabular}

\end{center}

\caption{\label{tab:QM}
  Effects of quadrupole--monopole terms, for $(10+1.4)M_{\odot}$
  binaries with maximally spinning BH.  At the beginning of each
  column we quote the overlaps between $\mathrm{ST}_2$ templates and
  $\mathrm{ST}_2^\mathrm{QM}$ templates that include
  quadrupole--monopole effects. Just as in Table
  \protect\ref{CauchyST123}, these overlaps are maximizing only over
  the extrinsic parameters $t_0$ and $\Phi_0$.  In parentheses,
  ``(...)'', we show the fitting factors for the
  $\mathrm{ST}_2^\mathrm{QM}$ family as matched by the $\mathrm{ST}_2$
  family; in brackets, ``[...]'', we show the intrinsic parameters at
  which the fitting factors are achieved.  The ``view'' column
  describes the position of the detector with respect to the initial
  orbital plane. In all cases the integration of the equations is
  started at a GW frequency of $60$ Hz.
}
\end{table*}

\section{A new physical template family for NS--BH and BH--BH precessing binaries}
\label{sec:parametrization}

In this section we discuss the detection of GWs from single-spin
precessing binaries using the template family first suggested in BCV2,
and further discussed in Sec.\ \ref{sec:template}. The proposed
detection scheme involves the deployment of a discrete template bank
along the relevant range of the intrinsic parameters $M$, $\eta$,
$\chi_1$, and $\kappa_1$, and the use of a detection statistic that
incorporates the maximization of the overlap over all the extrinsic
parameters: the directional angles $\Theta$, $\varphi$, $\theta$,
$\phi$, and $\psi$, the time of arrival $t_0$, and the initial phase
$\Phi_0$. In Sec.\ \ref{subsec:reparametrize} we describe the
reparametrization of the templates used for the formulation of the
maximized statistic, which is then discussed in Sec.\ 
\ref{subsec:maximize}, where we also present an approximated but
computationally cheaper version. The exact and
approximated statistics are discussed together in Sec.\ 
\ref{sec:testing} in the context of an optimized two-stage detection
scheme.

\subsection{Reparametrization of the waveforms}
\label{subsec:reparametrize}

We recall from Eqs.\ \eqref{generich}--\eqref{Fcross} that the generic
functional form of our precessing templates is
\begin{equation}
\label{eq:family}
h[\lambda^A] = Q^{ij}[M,\eta,\chi_1,\kappa_1;\Phi_0,t_0;t] P_{ij}[\Theta,\varphi;\theta,\phi,\psi].
\end{equation}
[Please note that for the rest of this paper we shall use coupled
raised and lowered indices to denote contraction; however, the
implicit metric is always Euclidian, so covariant and contravariant
components are equal. This will be true also for the STF components
introduced later, which are denoted by uppercase roman indices.]
  
The factor $Q^{ij}(t)$ (which describes the time-evolving dynamics of
the precessing binary) is given by
\beq
Q^{ij}=-\frac{2{\mu}}{D}\frac{M}{r} \Bigl[[\mathbf{e}_{+}]^{ij}
\cos2(\Phi+\Phi_{0})+[\mathbf{e}_{\times}]^{ij}
\sin2(\Phi+\Phi_{0})\Bigr], 
\eeq
where the GW phase $\Phi(t)$ and the GW polarization tensors $\mathbf{e}_{+,\times}(t)$ evolve according to the equations 
(\ref{eprecess}), (\ref{omegae}) and (\ref{epet}).
This factor depends on the intrinsic parameters $M$, $\eta$, $\chi_1$,
and $\kappa_1$, and on two extrinsic parameters: the initial phase
$\Phi_{0}$, and the time of arrival $t_0$ of the waveform, referred to
a fiducial GW frequency. 
We can factor out the initial phase $\Phi_0$ by defining
\bea
\label{Q0def}
Q^{ij}_{0} &\equiv& Q^{ij}(\Phi_{0}=0)\,, \\
\label{Q1def}
Q^{ij}_{\pi/2}&\equiv& Q^{ij}(\Phi_{0}=\pi/4)\,;
\eea
 we then have 
\begin{equation}
Q^{ij}=Q^{ij}_{0}\cos(2\Phi_{0})+Q^{ij}_{\pi/2}\sin(2\Phi_{0}).
\label{eq:simplpsi}
\end{equation}
The factor $P_{ij}$ (which describes the static relative position and
orientation of the detector with respect to the axes initially defined
by the binary) is given by
\begin{equation}
\label{Pijdef}
P_{ij}=[\mathbf{T}_{+}]_{ij} F_{+}+[\mathbf{T}_{\times}]_{ij} F_{\times},
\end{equation}
where the detector antenna patterns $F_{+,\times}(\theta,\phi,\psi)$
and the detector polarization tensors
$\mathbf{T}_{+,\times}(\Theta,\varphi)$ depend on the orientation
angles $\theta$, $\phi$, and $\psi$, and on the position angles
$\Theta$ and $\varphi$, all of them extrinsic parameters. The antenna
patterns can be rewritten as
\begin{equation}
\left\{
\begin{array}{c}
F_{+} \\
F_{\times}
\end{array}
\right\}
= \sqrt{F_{+}^{2}+F_{\times}^{2}}
\left\{
\begin{array}{c}
\cos \alpha \\
\sin \alpha
\end{array}
\right\};
\end{equation}
the factor $F \equiv \sqrt{F_{+}^{2}+F_{\times}^{2}}$ then enters $h$
as an overall multiplicative constant.\footnote{In fact, multiple GW
  detectors are needed to disentangle this factor from the distance
  $D$ to the source.} In what follows we shall be considering
normalized signals and templates, where $F$ drops out, so we set $F =
1$. We then have
\begin{equation}
P_{ij}=[\mathbf{T}_{+}]_{ij}\cos\alpha+[\mathbf{T}_{\times}]_{ij}\sin\alpha.
\label{eq:simplalpha}
\end{equation}

Both $Q^{ij}(t)$ and $P_{ij}$ are three-dimensional symmetric,
trace-free (STF) tensors, with five independent components each. Using
an orthonormal STF basis $M^{I}_{ij}$, $I = 1, \ldots, 5$, with 
$(M^I)_{ij} (M^J)^{ij} = \delta^{IJ}$,
we can conveniently express $P_{ij}$ and $Q^{ij}$ in terms of their components
on this basis,
\begin{equation}
Q^{ij} = Q^{I}(M^{I})^{ij}\,, \quad P_{ij} = P^{I}(M^{I})_{ij}\,,
\end{equation}
where
\begin{equation}
Q^{I} = Q^{ij}(M^{I})_{ij}\,, \quad P^{I} = P_{ij}(M^{I})^{ij}\,. 
\end{equation}

In this paper, we shall adopt a particular orthonormal basis,
\begin{eqnarray}
(M^1)_{ij}&=&\sqrt{\frac{4\pi}{15}}(\mathcal{Y}_{ij}^{22}+\mathcal{Y}_{ij}^{2-2}), \nonumber \\
(M^2)_{ij}&=&-i\sqrt{\frac{4\pi}{15}}(\mathcal{Y}_{ij}^{22}-\mathcal{Y}_{ij}^{2-2}), \nonumber \\
(M^3)_{ij}&=&-\sqrt{\frac{4\pi}{15}}(\mathcal{Y}_{ij}^{21}-\mathcal{Y}_{ij}^{2-1}), \\
(M^4)_{ij}&=&i\sqrt{\frac{4\pi}{15}}(\mathcal{Y}_{ij}^{21}+\mathcal{Y}_{ij}^{2-1}), \nonumber \\
(M^5)_{ij}&=&-\sqrt{\frac{8\pi}{15}}\mathcal{Y}_{ij}^{20}, \nonumber 
\end{eqnarray}
with $\mathcal{Y}^{2m}_{ij}$ defined by
\begin{equation}
\mathcal{Y}^{2m}_{ij}q^i q^j \equiv Y^{2m}(\hat{\mathbf{q}}),
\end{equation}
where $Y^{2m}(\hat{\mathbf{q}})$, $m=-2,\ldots,2$ are the usual $l=2$
spherical harmonics, and $\hat{\mathbf{q}}$ is any unit vector.  We bring
together this result with Eqs.\ \eqref{eq:simplpsi} and
\eqref{eq:simplalpha} to write the final expression
\begin{equation}
h = P_I
\Bigl( Q^{I}_{0}\cos(2\Phi_{0})+Q^{I}_{\pi/2}\sin(2\Phi_{0}) \Bigr),
\label{eq:finaltemplate}
\end{equation}
where
\begin{equation}
\label{PI}
P_I(\Theta,\varphi,\alpha) = \Bigl(
[\mathbf{T}_{+}(\Theta,\varphi)]_{I}\cos\alpha+[\mathbf{T}_{\times}(\Theta,\varphi)
]_{I}\sin\alpha \Bigr).
\end{equation}
Henceforth, we shall denote the surviving extrinsic
parameters collectively as $\Xi^{\alpha} \equiv (t_0, \Phi_0,
\alpha,\Theta,\varphi)$, and the intrinsic parameters as $X^i \equiv
(M,\eta,\chi_1,\kappa_1)$.

\newcommand{\rhocon}{{\rho_{\Xi^{\alpha}}}}
\newcommand{\rhounc}{{\rho'_{\Xi^{\alpha}}}}

\subsection{Maximization of the overlap over the extrinsic parameters}
\label{subsec:maximize}

As we have anticipated, it is possible to maximize the overlap $\rho =
\langle s,\hat{h}\rangle$ \emph{semialgebraically} over the extrinsic
directional parameters $\Theta$, $\varphi$, $\theta$, $\phi$, and
$\psi$, without computing the full representation of $\hat{h}$ for each of
their configurations. In addition, it is possible to maximize
efficiently also over $t_0$ and $\Phi_0$, which are routinely treated
as extrinsic parameters in nonspinning-binary GW searches.

For a given stretch of detector output $s$, and for a particular set of
template intrinsic parameters $X^i=(M, \eta, \chi_1, \kappa_1)$, we denote the
fully maximized overlap as
\begin{widetext}
\begin{equation}
\label{rhogeneric}
\rho_{\Xi^{\alpha}} \equiv \, \max_{\Xi^{\alpha}} \, \langle s,
\hat{h}(X^i,\Xi^{\alpha}) \rangle \, = \max_{t_0,\Phi_0,\Theta,\varphi,\alpha}
\underbrace{
\left[
\frac{P_I \left[\langle s , Q_0^I \rangle_{t_0}  \cos2\Phi_0 + \langle s ,
Q_{\pi/2}^I \rangle_{t_0}  \sin 2\Phi_0 \right]}{
\sqrt{P_I P_J \langle Q_0^I \cos 2\Phi_0 + Q_{\pi/2}^I \sin 2\Phi_0, 
Q_0^J \cos 2\Phi_0 + Q_{\pi/2}^J \sin 2\Phi_0 \rangle}}
\right]
}_{\displaystyle \rho}\,,
\end{equation}
\end{widetext}
where the subscript $t_0$ denotes the dependence of the
signal--template inner products on the time-of-arrival parameter of
the templates. In fact, each of these inner products can be computed
simultaneously for all $t_0$ with a single FFT; in this
sense, $t_0$ is an extrinsic parameter \cite{Schutz}.

Let us now see how to deal with $\Phi_0$. We start by making an
approximation that will be used throughout this paper.  We notice
that the template components $P_I Q_{0}^{I}$ and $P_I Q_{\pi/2}^{I}$
[Eqs.~\eqref{Q0def} and \eqref{Q1def}] are nearly orthogonal, and have
approximately the same signal power,
\bea
\label{q0q1}
\langle P_I Q_0^I, P_J Q_{\pi/2}^J \rangle &\simeq& 0\,, \\
\label{q0q0}
\langle P_I Q_0^I, P_J Q_0^J \rangle &\simeq&
\langle P_I Q_{\pi/2}^I, P_J Q_{\pi/2}^J \rangle\,;
\eea
this is accurate as long as the timescales for the
radiation-reaction--induced evolution of frequency and for the
precession-induced evolution of phase and amplitude modulations are
both much longer than the orbital period. More precisely, Eqs.\ 
\eqref{q0q1} and \eqref{q0q0} are valid up to the leading-order
stationary-phase approximation.  Under this hypothesis
Eq.~(\ref{rhogeneric}) simplifies, and its maximum over $\Phi_0$ is
found easily:
\bea
\label{rhoapp}
\rho_{\Xi^{\alpha}} &=& \max_{t_0,\Phi_0,\Theta,\varphi,\alpha}\frac{P_I
  \left[\langle s , Q_0^I \rangle \cos2\Phi_0 + \langle s ,
    Q_{\pi/2}^I \rangle \sin 2\Phi_0 \right] }{ \sqrt{P_I P_J \langle
    Q_0^I, Q_0^J\rangle }
} \nonumber \\
&=&\max_{t_0,\Theta,\varphi,\alpha} \sqrt{\frac{P_I P_J A^{IJ}}{P_I P_J
    B^{IJ}}}\equiv \max_{t_0,\Theta,\varphi,\alpha} \rho_{\Phi_0}
\,,
\eea
where we have defined the two matrices
\bea
A^{IJ} &\equiv& \langle s , Q_0^I \rangle_{t_0} \langle s , Q_0^J
    \rangle_{t_0} + \langle s , Q_{\pi/2}^I \rangle_{t_0} \langle s ,
    Q_{\pi/2}^J 
    \rangle_{t_0}  \,, \nonumber \\
B^{IJ} &\equiv& \langle Q_0^I , Q_0^J \rangle\,,
\eea
which are functions only of the intrinsic parameters (and, for
$A^{IJ}$, of $t_0$). We have tested the approximations \eqref{q0q1}
and \eqref{q0q0} by comparing the maximized overlaps obtained from
Eq.~\eqref{rhoapp} with the results of full numerical maximization
without approximations; both the values and the locations of the
maxima agree to one part in a thousand, even for systems with
substantial amplitude and phase modulations, where the approximations
are expected to be least accurate.

Although Eq.~\eqref{rhoapp} looks innocent enough, the maximization of
$\rho_{\Phi_0}$ is not a trivial operation. The five components of $P_I$ in
Eq.~\eqref{rhoapp} are not all independent, but they are specific
functions of only three parameters, $\Theta$, $\varphi$, and $\alpha$
[see the discussion leading to Eqs.~\eqref{eq:simplalpha} and
\eqref{PI}.] We can therefore think of $\rhocon$ as the result of
maximizing $\rho_{\Phi_0}$ with respect to the five-dimensional vector
$P_I$, \emph{constrained to the three-dimensional physical
submanifold} $P_I(\Theta,\varphi,\alpha)$. We shall then refer to
$\rhocon$ as the {\it constrained} maximized overlap.

What is the nature of the constraint surface? We can easily find the
two constraint equations that define it. First, we notice from Eqs.\
\eqref{rhogeneric} and \eqref{rhoapp} that the magnitude of the vector
$P_I$ does not affect the overlap: so we may rescale $P_I$ and set one
of the constraints as $P_I P^I = 1$; even better, we may require that
the denominator of Eq.\ \eqref{rhoapp} be unity, $P_I P_J B^{IJ}=1$.
Second, we remember that $P_{ij}$ [Eq.\ \eqref{Pijdef}] is the
polarization tensor for a plane GW propagating along the direction
vector
\begin{equation}
\label{eq:directionvector}
\hat{N}^i =
(\sin\Theta \cos\varphi,\,\sin\Theta\sin\varphi,\,\cos\Theta).
\end{equation}
Because GWs are transverse, $P_{ij}$ must admit $\hat{N}^i$ as an
eigenvector with null eigenvalue; it follows that
\beq
\label{detP}
\det P_{ij}=0.
\eeq
This equation can be turned into the second constraint for
the $P_I$ [see Eq.~\eqref{detPIJK} of App.~\ref{app:maximize}].

Armed with the two constraint equations, we can reformulate our
maximization problem using the method of Lagrangian multipliers
[Eq.~\eqref{eq:Lag} in App.~\ref{app:maximize}]. However, the
resulting system of cubic algebraic equations does not appear to have
closed-form analytic solutions. In App.~\ref{app:maximize} we develop
an iterative algebraic procedure to solve the system, obtaining the
constrained maximum and the corresponding $P_I$.  In practice, we have
found it operationally more robust to use a closed-form expression for
the partial maximum over $\Phi_0$ and $\alpha$ (which seems to be the
farthest we can go analytically), and then feed it into a numerical
maximum-finding routine (such as the well-known \texttt{amoeba}
\cite{nrc}) to explore the $(\Theta,\varphi)$ sphere, repeating this
procedure for all $t_0$ to obtain the full maximum.

To maximize $\rho_{\Phi_0}$ over $\alpha$, we use Eq.\ \eqref{PI} to
factor out the dependence of the $P_I$ on $\alpha$, and write
\beq
\label{alphaexp}
\sqrt{\frac{ P_I P_J A^{IJ}}{ P_I P_J B^{IJ}}}=\sqrt{\frac{ \mathbf{u}
    \mathbf{A}_\alpha \mathbf{u}^T }{\mathbf{u} \mathbf{B}_\alpha
    \mathbf{u}^T }}\,,
\eeq
where $\mathbf{u}$ is the two-dimensional row vector $(\cos\alpha,
\sin\alpha)$, and where $\mathbf{A}_{\alpha}$ and $\mathbf{B}_{\alpha}$
are the two-by-two matrices
\bea
\mathbf{A}_{\alpha} &=&
A^{IJ} 
\left(
\begin{array}{ll}
[\mathbf{T}_{+}]_{I}[\mathbf{T}_{+}]_{J} & 
[\mathbf{T}_{+}]_{I}[\mathbf{T}_{\times}]_{J} \\ %
{[\mathbf{T}_{+}]_{I}}[\mathbf{T}_{\times}]_{J} & 
[\mathbf{T}_{\times}]_{I}[\mathbf{T}_{\times}]_{J}
\end{array}
\right)\,,   \\
\mathbf{B}_{\alpha} &=&
B^{IJ}
\left( \begin{array}{lll}
[\mathbf{T}_{+}]_{I}[\mathbf{T}_{+}]_{J} & 
[\mathbf{T}_{+}]_{I}[\mathbf{T}_{\times}]_{J} \\ %
{[\mathbf{T}_{+}]_{I}}[\mathbf{T}_{\times}]_{J} & 
[\mathbf{T}_{\times}]_{I}[\mathbf{T}_{\times}]_{J} 
\end{array} \right)\,;
\eea
in these definitions we sum over the indices $I$ and $J$.  The
matrices $\mathbf{A}_{\alpha}$ and $\mathbf{B}_{\alpha}$ are
implicitly functions of the angles $\Theta$ and $\varphi$ through the
polarization tensors $\mathbf{T}_{+}$ and $\mathbf{T}_{\times}$.  It
is straightforward to maximize Eq.\ \eqref{alphaexp} over $\alpha$,
yielding\footnote{Just as it happens for the $P_I$, the magnitude of
$\mathbf{u}$ does not affect the value of Eq.~\eqref{alphaexp}, so the
maximization can be carried out equivalently over all the vectors
$\mathbf{u}$ that satisfy $\mathbf{u} \mathbf{B}_{\alpha}
\mathbf{u}^T=1$. We can then use a Lagrangian-multiplier method to
find the maximum, Eq.\ \eqref{rhoThetavarphi}, and the corresponding
$\mathbf{u}$, in a manner similar to the procedure used in
App.~\ref{app:maximize}.}
\beq
\label{rhoThetavarphi}
\rhocon = \max_{t_0,\Theta,\varphi} \sqrt{\max {\rm eigv}
  \left[\mathbf{A}_\alpha \,\mathbf{B}_{\alpha}^{-1}\right]}
\equiv \max_{t_0,\Theta,\varphi} \rho_{\Phi_0, \alpha} \, .
\eeq
The overlap $\rho_{\Phi_0, \alpha}$ is essentially equivalent to the
$\mathcal{F}$ statistic used in the search of GWs from pulsars
\cite{JKS98S2cont}.

The last step in obtaining $\rhocon$ is to maximize $\rho_{\Phi_0,
\alpha}$ numerically over the $(\Theta,\varphi)$ sphere, repeating
this procedure for all $t_0$ to obtain the full maximum. Now, $t_0$
enters Eq.\ \eqref{rhoThetavarphi} only through the ten
signal--template inner products $\langle s, Q_{0,\pi/2}^I \rangle$
contained in $\mathbf{A}_\alpha$, and each such product can be
computed for all $t_0$ with a single FFT. Even then, the semialgebraic
maximization procedure outlined above can still be very
computationally expensive if the search over $\Theta$ and $\varphi$
has to be performed for each individual $t_0$.  We have been able to
reduce computational costs further by identifying a rapidly computed,
fully algebraic statistic $\rhounc$ that approximates $\rhocon$ from
above. We then economize by performing the semialgebraic maximization procedure only
for the values of $t_0$ for which $\rhounc$ rises above a certain
threshold; furthermore, the location of the approximated maximum
provides good initial guesses for $\Theta$ and $\varphi$, needed to
kickstart their numerical maximization.

Quite simply, our fast approximation consists in neglecting the
functional dependence of the $P_I$ on the directional parameters,
computing instead the maximum of $\rho_{\Phi_0}$ 
[Eq.\ \eqref{rhoapp}] as if the five
$P_I$ were free parameters.  Using the method of
Lagrangian multipliers outlined in the beginning of 
App.~\ref{app:maximize} [Eqs.~\eqref{PIunconstrained}--\eqref{unc:app}], we get 
\begin{equation}
\label{rhounc}
\rho'_{\Xi^{\alpha}} = \max_{P_I} \sqrt{\frac{P_I P_J A^{IJ}}{P_I P_J B^{IJ}}}
= \sqrt{\max {\rm eigv} \left[\mathbf{A}\, \mathbf{B}^{-1}\right]}\,,
\end{equation}
with
\beq
(A^{IJ}-\lambda B^{IJ})P_J=0\,,\qquad  \lambda = \max {\rm eigv}
 [\mathbf{A}\,\mathbf{B}^{-1}]\,.
\eeq
Here the prime stands for {\it unconstrained} maximization over $P_I$.
We shall henceforth refer to $\rho'_{\Xi^{\alpha}}$ as the {\it
  unconstrained maximum}.

Note that the value of the $P_I$ at the unconstrained maximum will not in
general correspond to a physical set of directional parameters, so
$P_{ij}$ will not admit any direction vector $\hat{N}^i$ [Eq.\ 
\eqref{eq:directionvector}] as a null eigenvector. However, we can still get
approximate values of $\Theta$ and $\varphi$ by using instead the
eigenvector of $P_{ij}$ with the smallest eigenvalue (in absolute
value).


\section{Description and test of a two-stage search scheme}
\label{sec:testing}

In Sec.\ \ref{sec:parametrization} we have described a robust
computational procedure to find the maximum overlap $\rhocon$ (which
is maximized over the extrinsic parameters $\Phi_0$, $t_0$, and $P_I$,
where the allowed values of the $P_I$ are constrained by their
functional dependence on the directional angles).  We have also
established a convenient analytic approximation for $\rhocon$, the
unconstrained maximized overlap $\rhounc$ (which is maximized over the
extrinsic parameters $\Phi_0$, $t_0$, and $P_I$, but where the $P_I$
are treated as five independent and unconstrained coefficients).
Because the unconstrained maximization has access to a larger set of
$P_I$, it is clear that $\rhounc > \rhocon$. Still, at least when the
target signal $s$ is very close to the template $h(X_i)$, we expect
$\rhounc$ to be a very good approximation for $\rhocon$.

A quick look at the procedures outlined in Sec.\ 
\ref{sec:parametrization} shows that, for the filtering of
experimental data against a discrete bank of templates
$\{h(X_{(k)}^i)\}$, the computation of $\rhounc$ is going to be much
faster than the computation of $\rhocon$.  Under these conditions, it
makes sense to implement a two-stage search scheme where the discrete
bank $\{h(X_{(k)}^i)\}$ is first reduced by selecting the templates
that have high $\rhounc$ against the experimental data; at this stage
we identify also the promising times of arrival $t_0$. The exact
$\rhocon$ is computed only for these first-stage triggers, and
compared with the detection threshold $\rho^*$ to identify detection
candidates (one would use the same threshold $\rho^*$ in the first
stage to guarantee that all the detection candidates will make it into
the second stage).\footnote{This is not a conventional hierarchical
  scheme, at least not in the sense that there is a tradeoff between
  performance and accuracy.}

To prove the viability of such a search scheme, we shall first
establish that $\rhounc$ is a good approximation for $\rhocon$ for
target signals and templates computed using the adiabatic model of
Sec.\ \ref{sec:template}. We will take slightly displaced intrinsic
parameters for target signals and templates, to reproduce the
experimental situation where we are trying to detect a signal of
arbitrary physical parameters with the closest template belonging to a
discrete bank. This first test is described in Sec.\
\ref{sec:testone}. We shall then study the false-alarm statistics of
$\rhocon$ and $\rhounc$, and we shall show that, for a given detection
threshold, the number of first-stage triggers caused by pure noise is
only a few times larger than the number of \emph{bona fide}
second-stage false alarms.  Such a condition is necessary because the
two-stage detection scheme is computationally efficient only if few
templates need ever be examined in the expensive second stage. The
false-alarm statistics (in Gaussian stationary noise) are obtained in
Sec.\ \ref{sec:falsealarm}, and the second test is described in Sec.\
\ref{sec:testtwo}.

\subsection{Numerical comparison of constrained and unconstrained
maximized overlaps}
\label{sec:testone}

In this section we describe a set of Monte Carlo runs designed to test
how well $\rhounc$ can approximate $\rhocon$, for the target signals
and templates computed using the adiabatic model of Sec.\ 
\ref{sec:template}, for typical signal parameters, and for
signal--template parameter displacements characteristic of an actual
search.

We choose target signals with 20 different sets of intrinsic
parameters given by
\begin{equation}
(m_1, m_2, \chi_1, \kappa_1) = 
\left\{
\begin{array}{c}
(10,1.4)M_\odot \\
(7,3)M_\odot
\end{array}
\right\}
\times
\left\{
\begin{array}{c}
0.5 \\
1
\end{array}
\right\} \times
\left\{
\begin{array}{c}
-0.9 \\ -0.5 \\ 0.0 \\ 0.5 \\ 0.9
\end{array}
\right\}.
\end{equation}
For each set of target-signal intrinsic parameters, we choose 100
random sets of extrinsic parameters $(\Theta,\varphi,\alpha,\Phi_0)$,
where the combination $(\Theta,\varphi)$ is distributed uniformly on
the solid angle, and where $\alpha$ and $\Phi_0$ are distributed
uniformly in the $[0,2\pi]$ interval. The target signals are
normalized, so the allowed range for $\rhocon$ and $\rhounc$ is
$[0,1]$.

For each target signal, we test 50 (normalized) templates displaced in
the intrinsic-parameter space $(M,\eta,\chi_1,\kappa_1)$ [the optimal
extrinsic parameters will be determined by the optimization of
$\rhocon$ and $\rhounc$, so we do not need to set them]. The
direction of the displacement is chosen randomly in the
$(M,\eta,\chi_1,\kappa_1)$ space. For simplicity, the magnitude of the displacement is chosen so that, for each set of target-signal
intrinsic parameters and for the \emph{first set} of target-signal
extrinsic parameters, the overlap $\rhounc$ is about 0.95; the
magnitude is then kept fixed for the other 99 extrinsic-parameter
sets, so $\rhounc$ can be very different in those cases.
\begin{figure*}
\begin{center}
\includegraphics{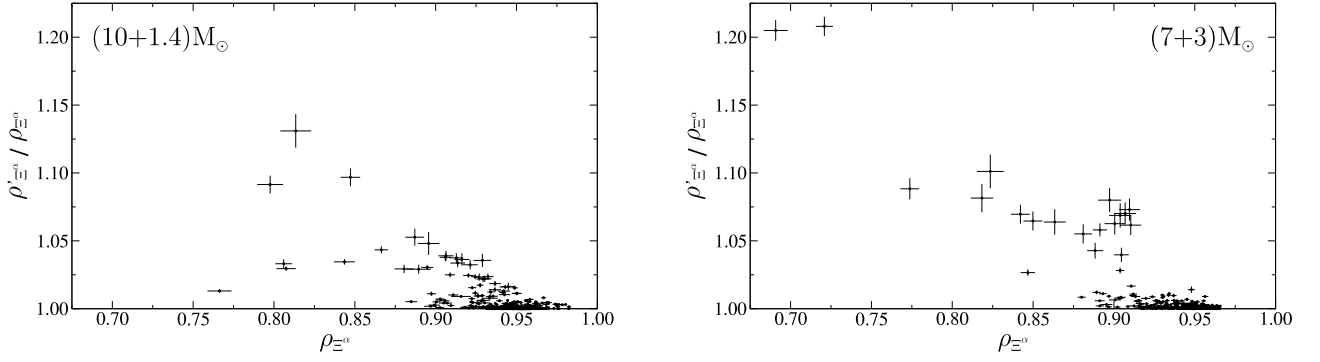}
\caption{\label{ratiovscon}
  Ratio between the unconstrained ($\rhounc$) and constrained
  ($\rhocon$) maximized overlaps, as a function of $\rhocon$. Each
  point corresponds to one out of $20 \times 50$ sets of intrinsic
  parameters for target signal and template, and is averaged over 100
  sets of extrinsic parameters for the target signal. The error bars
  show the standard deviations of the sample means (the standard
  deviations of the samples themselves will be 10 times larger, since
  we sample 100 sets of extrinsic parameters).  The two panels show
  results separately for $(10+1.4)M_\odot$ (left) and $(7+3)M_\odot$
  target systems (right).  The few points scattered toward higher
  ratios and lower $\rhocon$ are obtained when the first set of
  extrinsic parameters happens to yield a high $\rhounc$ that is not
  representative of most other values of the extrinsic parameters;
  then the magnitude of the intrinsic-parameter deviation is set too
  high, and the comparison between $\rhounc$ and $\rhocon$ is done at
  low $\rhocon$, where the unconstrained maximized overlap is a poor
  approximation for its constrained version.}
\end{center}
\end{figure*}

Figure \ref{ratiovscon} shows the ratio $\rhounc/\rhocon$, for
each pair [$20 \times 50$ in total] of target and template
intrinsic-parameter points, averaged over the 100 target
extrinsic-parameter points, as a function of the averaged
$\rhocon$. The $\rhounc$ get closer to the $\rhocon$ as the
latter get higher; most important, the difference is within $\sim$ 2\%
when $\rhocon > 0.95$, which one would almost certainly want to
achieve in an actual search for signals.  We conclude that
$\rhounc$ can indeed be used as an approximation for $\rhocon$
in the first stage of a two-stage search. The second stage is still
necessary, because the false-alarm statistics are worse for the
unconstrained maximized overlap (where more degrees of freedom are
available) than for its constrained version. We will come back to this
in the next two sections.
\begin{figure*}
\begin{center}
\includegraphics{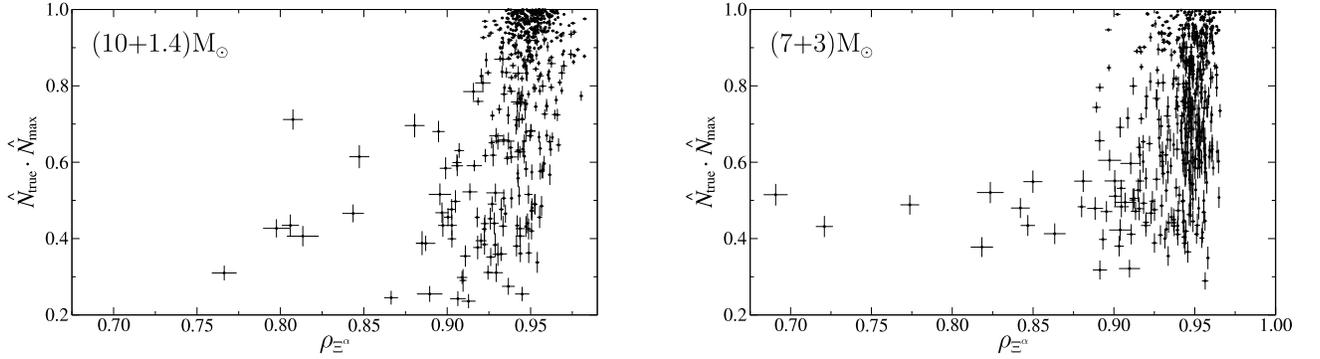}
\caption{\label{ndotn0}Inner product between target-signal source
direction $\hN_\mathrm{true}$ and $\rhocon$-maximizing
source direction $\hN_\mathrm{max}$, as a function of
$\rhocon$. Each point corresponds to one out of $20 \times 50$ sets
of intrinsic parameters for target signal and template, and is
averaged over 100 sets of extrinsic parameters for the target
signal. Standard deviations of the sample means are
shown as error bars, as in Figure \ref{ratiovscon}.
The two panels show separately $(10+1.4)M_\odot$ target systems (left) and
$(7+3)M_\odot$ target systems (right).}
\end{center}
\end{figure*}

It is also interesting to compare the set of extrinsic parameters of
the target signal with the set of extrinsic parameters that maximize
$\rhocon$, as characterized by the corresponding source direction
vectors, $\hN_\mathrm{true}$ and $\hN_\mathrm{max}$ respectively.
Figure \ref{ndotn0} shows the inner product $\hN_\mathrm{true} \cdot
\hN_\mathrm{max}$, averaged over the 100 target extrinsic-parameter
points, as a function of the averaged $\rhocon$.  The difference
between the vectors can be very large, even when $\rhocon > 0.95$:
this happens because the intrinsic-parameter displacement between
target signal and template can be compensated by a change in the
extrinsic parameters of template (in other words, the effects of the
intrinsic and extrinsic parameters on the waveforms are highly
correlated).

\subsection{False-alarm statistics for the constrained and unconstrained maximized overlaps}
\label{sec:falsealarm}

In this section we derive and compare the false-alarm statistics of
$\rhocon$ and $\rhounc$. Our purpose is to estimate the number of
additional triggers that are caused by replacing the detection
statistic $\rhocon$ by the first-stage statistic $\rhounc$. Our
two-stage detection scheme, which employs the rapidly computed
$\rhounc$ to choose candidates for the more computationally expensive
$\rhocon$, will be viable only if the number of those candidates is
small enough.

By definition, a false alarm happens when, with interferometer output
consisting of pure noise, the detection statistic computed for a given
template happens to rise above the detection threshold. Although the
detection statistics $\rhocon$ and $\rhounc$ include maximization over
the time of arrival $t_0$, we find it convenient to exclude $t_0$ from
this computation, and to include it later when we evaluate the total
false-alarm probability for all the templates in the bank.

Recall that $\rhocon$ [Eq.~\eqref{rhoapp}] and $\rhounc$
[Eq.~\eqref{rhounc}] are functions of the matrices $\mathbf{A}$ and
$\mathbf{B}$, which contain the inner products $\langle s,
Q_{0,\pi/2}^I \rangle$ and $\langle Q_{0,\pi/2}^I, Q_{0,\pi/2}^J
\rangle$, respectively. In this case the signal $s$ is a realization
of the noise, $n$. We combine the vectors $Q_{0}^{I}$ and
$Q_{\pi/2}^{I}$ together as $Q^{\mathcal{I}}$ with $\mathcal{I} = 1,
\ldots, 10$; under the assumption of Gaussian stationary noise,
$Y^\mathcal{I} \equiv \langle n, Q^{\mathcal{I}}\rangle$ is a
ten-dimensional Gaussian random vector with zero mean and covariance
matrix \cite{Wainstein}
\begin{equation}
\label{Cijdef}
C^{\mathcal{IJ}}=\overline{\langle n,Q^{\mathcal{I}}\rangle\langle n, Q^{\mathcal{J}}\rangle}=
\langle Q^{\mathcal{I}},Q^{\mathcal{J}} \rangle\,.
\end{equation}
The covariance matrix $C^{\mathcal{IJ}}$ specifies completely the
statistical properties of the random vector $Y^\mathcal{I}$, and it is
a function only of $\mathbf{B}$, and therefore only of the intrinsic
parameters of the template. We can also combine
$P_I\cos 2\Phi_0$ and $P_I\sin 2\Phi_0$ together as $P_{\mathcal{I}}$,
and then write the maximized overlaps $\rho_{\Xi^{\alpha}}$ and
$\rho'_{\Xi^{\alpha}}$ as
\begin{equation}
\label{rho10}
\max_{P_{\mathcal{I}}} \frac{P_{\mathcal{I}} \langle
  n,Q^{\mathcal{I}}\rangle}{\sqrt{ P_{\mathcal{I}} P_{\mathcal{J}}
    \langle Q^\mathcal{I},Q^\mathcal{J}\rangle }} = \max_{P_{\mathcal{I}}}
\frac{P_{\mathcal{I}} Y^\mathcal{I}}{\sqrt{ P_{\mathcal{I}} P_{\mathcal{J}}
    C^\mathcal{IJ} }}\,,
\end{equation}
where maximization if performed over the appropriate range of the
$P_\mathcal{I}$. In the rest of this section
we shall use the shorthand $\rho$ to denote both $\rhocon$ and
$\rhounc$.

Equation\ \eqref{rho10} is very general: it describes $\rhocon$ and
$\rhounc$, but it can also incorporate other maximization ranges over
the $P_\mathcal{I}$, and it can even treat different template
families.  In fact, the maximized detection statistic for the $(\psi_0
\psi_{3/2} \mathcal{B})_6$ DTF of BCV2 can be put into the same form,
with $P_{\mathcal{I}} \equiv \alpha_\mathcal{I}$, for $\mathcal{I} =
1,\ldots,6$, and with completely unconstrained maximization.

We can now generate a distribution of the detection statistic $\rho$ for a
given set of intrinsic parameters by generating a distribution of the
Gaussian random vector $Y^{\mathcal{I}}$, and then computing $\rho$
from Eq.\ \eqref{rho10}. The first step is performed easily by
starting from ten independent Gaussian random variables
$Z^\mathcal{I}$ of zero mean and unit variance, and then setting
$Y^\mathcal{I} = [\sqrt{C}]^\mathcal{IJ} Z_\mathcal{J}$.\footnote{The
  square root of the matrix $[\sqrt{C}]^\mathcal{IJ}$ can be defined,
  for instance, by $\sqrt{\mathbf{C}} \sqrt{\mathbf{C}}^T =
  \mathbf{C}$, and it can always be found because the covariance
  matrix $C^\mathcal{IJ}$ is positive definite.  It follows that
  $\overline{Y^\mathcal{I} Y^\mathcal{J}} = [\sqrt{C}]^\mathcal{IL}
  [\sqrt{C}]^\mathcal{JM} \overline{X_\mathcal{L} X_\mathcal{M}} =
  [\sqrt{C}]^\mathcal{IL} [\sqrt{C}]^\mathcal{JM} \delta_\mathcal{LM}
  = C^\mathcal{IJ}$, as required.} Thus, there is no need to generate
actual realizations of the noise as time series, and no need to
compute the inner products $\langle n, Q^\mathcal{I} \rangle$ explicitly.

The statistics $\rho$ [Eq.\ (\ref{rho10})] are
homogeneous with respect to the vector $Z^{\mathcal{I}}$: that is, if
we define $Z^{\mathcal{I}} = r \hat{Z}^{\mathcal{I}}$ (where
$r\equiv\sqrt{Z^\mathcal{I} Z_\mathcal{I}}$ and $\hat{Z}^{\mathcal{I}}
\hat{Z}_\mathcal{I} = 1$) we have
\begin{equation}
\label{rhodecomp}
\rho[ Y^\mathcal{I}(Z^\mathcal{I}) ] =
r \rho[ Y^\mathcal{I}(\hat{Z}^\mathcal{I}) ] \equiv r \rho_1(\Omega)\,;
\end{equation}
here $\Omega$ represents the direction of $\hat{Z}^\mathcal{I}$ in its
ten-dimensional Euclidian space.
The random variable $r$ has the marginal probability density
\begin{equation}
p_r(r)=\frac{r^{\nu-1}\exp(-r^{2}/2)}{2^{\nu/2-1}\Gamma(\nu/2)}\,,
\end{equation}
where the direction $\Omega$ is distributed uniformly over a
ten-sphere.  [For the rest of this section we shall write equations in
the general $\nu$-dimensional case; the special case of our template
family is recovered by setting $\nu = 10$.] The random variables $r$
and $\Omega$ [and therefore $\rho_1(\Omega)$] are statistically
independent, so the cumulative distribution function for the statistic
$\rho$ is given by the integral
\begin{multline}
\label{eq:FA}
P(\rho<\rho^*) = \int \!\! d\Omega \int_0^{\rho^*/\rho_1(\Omega)}
p_r(r) \, dr \Big/ \!\! \int \!\! d\Omega \\
= 1 - \int \frac{
\Gamma\Bigl[\frac{\nu}{2},\frac{\bigl(\rho^*/\rho_1(\Omega)\bigr)^2}{2}\Bigr]
}{ \Gamma\bigl[\frac{\nu}{2}\bigr] }
d\Omega
\Big/ \!\! \int \!\! d\Omega \,,
\end{multline}
where $\Gamma[k,z] = \int_z^{+\infty}
t^{k-1} e^{-t} dt$ is the \emph{incomplete gamma function}.

The false-alarm probability for a single set of intrinsic parameters
and for a single time of arrival is then $1-P(\rho<\rho^*)$. The final
integral over the $\nu$-dimensional solid angle can be performed by
Monte Carlo integration, averaging the integrand over randomly chosen
directions $\Omega$. Each sample of the integrand is obtained by
generating a normalized $\hat{Z}^\mathcal{I}$ (that is, a direction
$\Omega$), obtaining the corresponding $Y^\mathcal{I}$, computing
$\rho_1(\Omega)$ from Eq.\ \eqref{rho10}, and finally plugging
$\rho_1(\Omega)$ into the $\Gamma$ function.

Equation \eqref{eq:FA} shows that if we set $\rho_1(\Omega)=1$, the
random variable $\rho$ follows the $\chi_{(\nu)}$ distribution; this
is obvious because in that case $\rho = r = \sqrt{Z^{\mathcal{I}}
  Z_{\mathcal{I}}}$ [see Eq.~\eqref{rhodecomp}], where the
$Z^\mathcal{I}$ are $\nu$ independent Gaussian random variables.  In
fact, $\rho_1(\Omega)$ can be written as
\begin{equation}
\rho_1(\Omega) = \max_{P^\mathcal{I}}
\frac{R_\mathcal{I} \hat{Z}^\mathcal{I}}{
\sqrt{R_\mathcal{I} R_\mathcal{J} \delta^\mathcal{IJ}}},
\quad \mathrm{where} \; R_\mathcal{I} = [\sqrt{C}]_\mathcal{IJ} P^\mathcal{J};
\end{equation}
which shows that $\rho_1(\Omega)=1$ uniformly for every $\Omega$ if
and only if the range of maximization for $P_{\mathcal{I}}$ is the
\emph{entire} $\nu$-dimensional linear space generated by the basis
$\{Q^{\mathcal{I}}\}$; however, once we start using the entire linear
space, the particular basis used to generate it ceases to be
important, so the covariance matrix $C^\mathcal{IJ}$ drops out of the
equations for the false-alarm probabilities.  That is the case, for
instance, for the $(\psi_0 \psi_{3/2} \mathcal{B})_6$ DTF [see Sec.\ V
B of BCV2], whose false-alarm probability is described by the
$\chi_{(\nu = 6)}$ distribution.  For our template family $\nu=10$,
but both $\rhocon$ and $\rhounc$ have very restrictive maximization
ranges for $P_{\mathcal{I}}$ (because $P_{\mathcal{I}=1,\ldots,5}$ and
$P_{\mathcal{I}=6,\ldots,10}$ are strongly connected): so both
$\rhocon$ and $\rhounc$ will have much lower false-alarm probability,
for the same threshold $\rho^*$, than suggested by the
$\chi_{(\nu=10)}$ distribution. In fact, in the next section we shall
see that the effective $\nu$ for the detection statistic $\rhounc$ is
about 6; while the effective $\nu$ for $\rhocon$ is even lower.

\subsection{Numerical investigation of false-alarm statistics}
\label{sec:testtwo}

The total false-alarm probability for the filtering of experimental
data by a template bank over a time $T$ is
\begin{equation}
P_\mathrm{tot}(\rho > \rho^*) = 1 - \bigl[P(\rho<\rho^*)\bigr]^{\mathcal{N}_\mathrm{shapes} \mathcal{N}_\mathrm{times}}
\end{equation}
(see for instance BCV1), where the exponent
$\mathcal{N}_\mathrm{shapes} \mathcal{N}_\mathrm{times}$ is an
estimate of the number of effective independent statistical tests. The
number of independent signal shapes $\mathcal{N}_\mathrm{shapes}$ is
related to (and smaller than) the number of templates in the
bank;\footnote{It is given, very roughly, by the number of templates
when the minimum match is set to 0. See for instance BCV1.} the number
of independent times of arrival $\mathcal{N}_\mathrm{times}$ is
roughly $T/\delta t_0$, where $\delta t_0$ is the mismatch in the time
of arrival needed for two nearby templates to have, on average, very
small overlap.  In our tests we set $\mathcal{N}_\mathrm{shapes} =
10^6$ and $\mathcal{N}_\mathrm{times}=3\times 10^{10}$ (or
equivalently $\delta t_0 \simeq 1$ ms), as suggested by the results of
Sec.\ \ref{sec:metric} for template counts and for the full mismatch
metric; in fact, both numbers represent rather conservative choices.

\begin{figure}
\vspace{0.5cm}
\begin{center}
\includegraphics{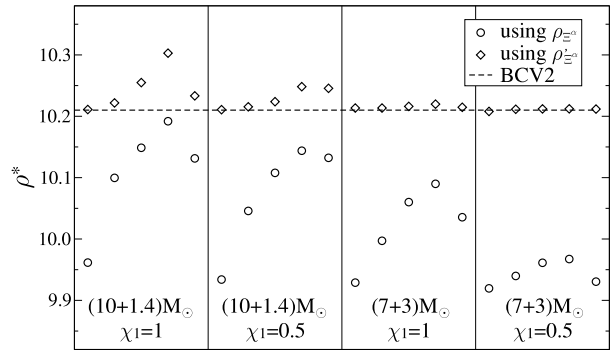}
\caption{\label{fas}Detection thresholds $\rho^*$ for a false-alarm
  rate of $10^{-3}$/year, using the constrained statistic $\rhocon$
  (circles), the approximated (unconstrained) statistic $\rhounc$
  (diamonds), and the detection statistic for the $(\psi_0 \psi_{3/2}
  \mathcal{B})_6$ DTF from BCV2 (dashed line). The four windows correspond
  to the masses and $\chi_1$ shown; the points in each window
  correspond to $\kappa_1 = 0.9$, $0.5$, $0$, $-0.5$, $-0.9$.}
\end{center}
\end{figure}

We compute single-test false-alarm probabilities from Eq.\ 
\eqref{eq:FA}, averaging the integrand over $10^5$ randomly chosen
values of $\Omega$ to perform the integration over $\Omega$, {\it \`a
  la} Monte Carlo.  Our convergence tests indicate that this many
samples are enough to obtain the required precision.\footnote{In fact,
  the average is dominated by the samples that yield the larger values
  of $\rho_1(\Omega)$, since the $\Gamma$ function amplifies small
  changes in its argument.  So the number of samples used in the Monte
  Carlo integration needs to be such that enough large
  $\rho_1(\Omega)$ do come up.} In Fig.~\ref{fas} we show the
thresholds $\rho^*$ required to achieve a total false-alarm rate of
$10^{-3}$/year; the figure suggests that a threshold close to 10 is
adequate. The thresholds are only marginally higher for the
unconstrained statistic, so the number of first-stage false alarms
that are dismissed in the second stage is limited. We show also the
threshold required to achieve the same false-alarm rate with the
$(\psi_0 \psi_{3/2} \mathcal{B})_6$ DTF of BCV2: this threshold is
very close to the values found for $\rhounc$, indicating that
$\rhounc$ has roughly six effective degrees of freedom (as it seems
reasonable from counting the five $P^I$ plus $\Phi_0$).  The BCV2
threshold is consistently higher than the $\rhocon$ threshold for the
same single-test false-alarm rate; this suggests that the detection
scheme discussed in this paper is less wasteful (with respect to the
available signal power) than the BCV2 scheme, assuming of course that
the number of templates used in the two banks is similar.
\begin{figure*}
\begin{center}
\includegraphics{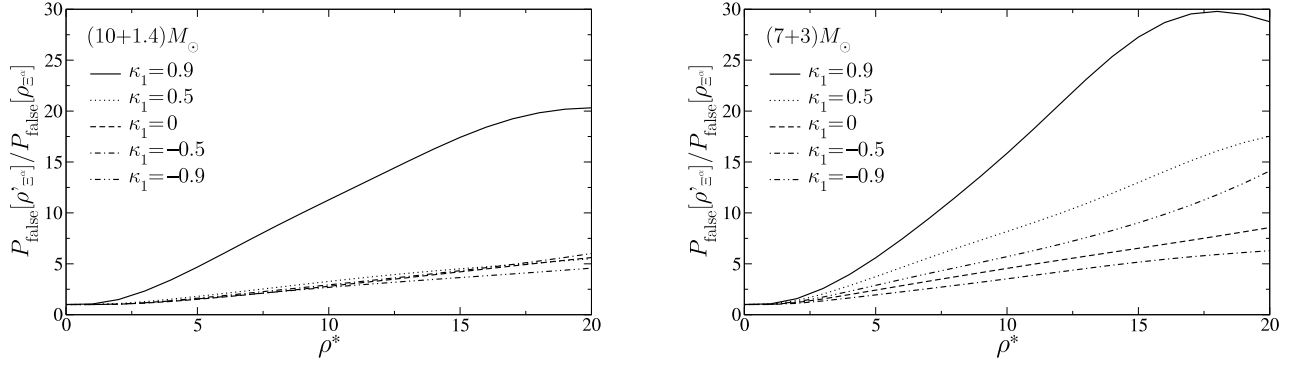}
\caption{\label{ratiovsthresh}Ratio
  $(1-P[\rhounc<\rho^*])/(1-P[\rhocon<\rho^*])$ between single-test
  false-alarm probabilities for the unconstrained and constrained
  detection statistics, as a function of threshold $\rho^*$. The two
  panels represent systems with masses equal to (10+1.4)$M_{\odot}$
  (left) and to (7+3)$M_{\odot}$ (right). The five curves in each plot
  correspond to different $\kappa_1$.}
\end{center}
\end{figure*}

In Fig.\ \ref{ratiovsthresh} we show the ratio between the single-test
false-alarm probabilities for $\rhocon$ and $\rhounc$: for a common
threshold around 10, we can expect about five times more false alarms
using $\rhounc$ than using $\rhocon$, for most values of the intrinsic
parameters (for some of them, this number could be as high as $\sim
15$).  These results corroborate our suggestion of using $\rhounc$ in
the first-stage of a two-stage detection scheme, to weed out most of
the detection candidates before computing the more computationally
expensive $\rhocon$.



\newcommand{\hh}{\hat{h}}

\section{Template counting and placement}
\label{sec:metric}

The last aspect to examine before we can recommend the template family
of Sec.\ \ref{sec:parametrization} for actual use with the two-stage
search scheme of Sec.\ \ref{sec:testing} is the total number of
templates that are needed in practice.  As mentioned in Sec.\
\ref{sec:refresher}, the template-bank size and geometry required to
achieve a certain minimum match can be studied using the \emph{mismatch
metric} \cite{BSD,O,OS99}, which describes, to quadratic order, the
degrading overlap between nearby elements in a template bank:
\begin{multline}
\label{eq:basicmetric}
1 - \langle \hh (\lambda^A), \hh (\lambda^A + \Delta \lambda^A)
\rangle \\
\equiv \delta[\lambda^A, \lambda^A + \Delta \lambda^A] =
g_{BC} \Delta \lambda^B \Delta \lambda^C, 
\end{multline}
where $\delta$ denotes the mismatch, and where
\begin{equation}
\label{eq:defmetric}
g_{BC} = -\frac{1}{2} \frac{\partial^2 \bigl\langle \hh (\lambda^A),
  \hh (\lambda^A + \Delta \lambda^A) \bigr\rangle}{\partial (\Delta \lambda^B) \partial (\Delta \lambda^C)}.
\end{equation}
No zeroth- or first-order terms are needed in the
expansion~\eqref{eq:basicmetric}, because the overlap has a maximum of $1$
(for normalized templates) at $\Delta \lambda^A = 0$. The metric is
positive definite, because $\delta > 0$. Note that, according to this
definition, the mismatch $\delta$ is the \emph{square} of the
\emph{metric distance} between $\lambda^A$ and $\lambda^A + \Delta
\lambda^A$. It is also {\it half} the square of the \emph{inner-product
distance} $\sqrt{\langle \Delta \hh, \Delta \hh\rangle}$, where
$\Delta \hh \equiv \hh(\lambda^A) - \hh(\lambda^A+\Delta
\lambda^A)$.\footnote{So $g_{BC}$ is truly a metric, except for a
factor of $1/2$.}

Ideally, for a given continuous template family, one could find a
reparametrization in which the metric is a Kronecker delta, and then
lay down a template bank as a uniform hypercubic lattice in these
coordinates, with the appropriate density to yield the required
MM. For a hypercubic lattice in $n$ dimensions,\footnote{For specific
dimensionalities, other regular packings might be more efficient: for
instance, in two dimensions a lattice of equilateral triangles
requires fewer templates than a lattice of squares.} the (metric)
side $\delta l$ of the lattice cell is given by the relation $1 -
\mathrm{MM} = n (\delta l/2)^2$ \cite{O,bcv1};
we then get the total number of templates in the bank by dividing the
total (metric) volume of parameter space by the volume of each cell:
\begin{equation}
\label{calN}
{\cal N}_\mathrm{templates}=
\int \! \sqrt{|\det g_{BC}|} d^n\lambda^A {\Big /}
\left[2 \sqrt{(1-{\rm MM})/n}\right]^n.
\end{equation}
In practice, this expression will usually underestimate the total
number of templates, for two reasons: first, for more than two
dimensions it is usually impossible to find coordinates where the
metric is diagonalized everywhere at once; second, the fact that the
actual parameter space is bounded will also introduce corrections to
Eq.\ \eqref{calN}. [The presence of null parameter directions, discussed
in Sec.\ \ref{sec:reduction}, can also be seen as an extreme case of
boundary effects.]

As we showed in Secs.\ \ref{sec:parametrization} and
\ref{sec:testing}, the overlap of the detector output with one of the
$\mathrm{ST}_N$ templates can be maximized automatically over all the
extrinsic parameters $\Xi^\alpha$; it follows that a discrete template
bank will need to extend only along the four intrinsic parameters
$X^i$. So the estimate \eqref{calN} for the number of templates should
be computed on the \emph{projected} metric $g^\mathrm{proj}_{ij}$ that
satisfies
\bea
\label{eq:gproj}
&&1-\rho_{\Xi^\alpha}\!\left[\hh(X^i,\Xi^\alpha),\hh(X^i+\Delta
  X^i)\right] \nonumber \\
&\equiv&
1-\max_{\Xi'^\alpha} \langle 
\hh(X^i,\Xi^\alpha),\hh(X^i+\Delta X^i,\Xi'^\alpha)\rangle \\
&
= &g^\mathrm{proj}_{ij} \Delta X^i \Delta X^j. \nonumber
\eea
Note that $g^\mathrm{proj}_{ij}$ is still a function of {\it all} the
parameters. In Sec.\ \ref{sec:projection} we compute
$g^\mathrm{proj}_{ij}$ from the full metric $g_{BC}$; we then proceed
to construct an {\it average} metric,
$\overline{g^\mathrm{proj}_{ij}}$, which is connected closely to
detection rates, and does not depend on the extrinsic parameters.

In fact, it turns out that \emph{not all four intrinsic parameters are
needed to set up a template bank that achieves a reasonable} MM: we
can do almost as well by replacing a 4-D bank with a 3-D bank where
(for instance) we set $\kappa_1 = 0$. As a geometrical counterpart to
this fact, the projected metric must allow a quasinull direction: that
is, it must be possible to move along a certain direction in parameter
space while accumulating almost no mismatch. The correct template
counting for the 3-D bank is then described by a \emph{reduced}
metric, which we discuss in Sec.\ \ref{sec:reduction}. Finally, we
give our results for the total number of templates in Sec.\
\ref{sec:counting}.

\subsection{Computation of the full, projected and average metric}
\label{sec:projection}

According to Eq.\ \eqref{eq:defmetric}, the full metric $g_{BC}$ can
be computed numerically by fitting the quadratic decrease of the
overlap $\langle \hh (\lambda^A), \hh (\lambda^A+\Delta \lambda^A) \rangle$
around $\Delta \lambda^A = 0$. It is also possible to rewrite $g_{BC}$
in terms of first-order derivatives of the waveforms: since $\langle
\hh(\lambda^A), \hh(\lambda^A) \rangle = 1$ for all $\lambda^A$,
\begin{equation}
\frac{\partial}{\partial \lambda^B} \bigl\langle \hh, \hh \bigr\rangle = 
2 \biggl\langle \hh, \frac{\partial \hh}{\partial \lambda^B} \biggr\rangle=0
\end{equation}
[in this equation and in the following, we omit the parametric
dependence $\hh(\lambda^A)$ for ease of notation]; taking one more
derivative with respect to $\lambda^C$, we get
\begin{equation}
\biggl\langle{\frac{\partial \hh}{\partial \lambda^C}},{\frac{\partial \hh}{\partial \lambda^B}}\biggr\rangle
+ \biggl\langle \hh, {\frac{\partial^2 \hh}{\partial \lambda^C\partial \lambda^B}}\biggr\rangle = 0,
\end{equation}
which implies [by Eq.~(\ref{eq:defmetric})]
\begin{equation}
\label{gfisher}
g_{BC} = \frac{1}{2}
\biggl\langle{\frac{\partial \hh}{\partial \lambda^B}},
{\frac{\partial \hh}{\partial \lambda^C}}\biggr\rangle.
\end{equation}
The inner product in the right-hand side of Eq.~\eqref{gfisher}
expresses the \emph{Fisher information matrix} for the normalized
waveforms $\hh (\lambda^A)$ (see for instance Ref.~\cite{Finn}); for
nonnormalized waveforms $h(\lambda^A)$ we can write
\begin{equation}
\begin{split}
g_{BC} = \quad & \frac{1}{2 \langle h, h \rangle}
\left\langle \frac{\partial h}{\partial \lambda^B}, \frac{\partial h}{\partial \lambda^C} \right\rangle \\
- \, & \frac{1}{2 \langle h, h \rangle^2}
\left\langle \frac{\partial h}{\partial \lambda^B}, h \right\rangle
\left\langle h, \frac{\partial h}{\partial \lambda^C}  \right\rangle\,.
\end{split}
\end{equation}
It is much easier to compute the mismatch metric from Eq.~\eqref{gfisher} 
rather than from Eq.\ \eqref{eq:defmetric}, for two reasons. First, 
we know the analytic dependence of the templates on all the extrinsic
parameters (except $t_0$), so we can compute the derivatives $\partial
\hh / \partial \Xi^\alpha$ analytically (the derivative with respect to
$t_0$ can be handled by means of the Fourier-transform time-shift
property $\mathcal{F}[h(t+t_0)] = \mathcal{F}[h(t)] \exp [2\pi i f
t_0]$). Second, although the derivatives $\partial \hh / \partial X^i$
have to be computed numerically with finite-difference expressions
such as $[\hh(X^i+\Delta X^i,\Xi^a) - \hh(X^i,\Xi^a)]/\Delta X^i$,
this is still easier than fitting the second-order derivatives of the
mismatch numerically.\footnote{We have found that we can obtain a
satisfactory precision for our metrics by taking several cautions: (i)
reducing the parameter displacement $\Delta X^i$ along a sequence
${}^{(k)}\!\Delta X^i$ until the norm $\langle {}^{(k)}\!\Delta \hh -
{}^{(k-1)}\!\Delta
\hh, {}^{(k)}\!\Delta \hh - {}^{(k-1)}\!\Delta \hh \rangle$ of the
$k$th correction becomes smaller than a certain tolerance, where
${}^{(k)}\!\Delta \hh = \left[\hh(X^i+{}^{(k)}\!\Delta X^i,\Xi^a) -
\hh(X^i,\Xi^a)\right]/{}^{(k)}\!\Delta X^i$ is the $k$th approximation
to the numerical derivative; (ii) employing higher-order finite-difference
expressions; (iii) aligning both the starting and ending times of the
waveforms $\hh(X^i,\Xi^a)$ and $\hh(X^i+\Delta X^i,\Xi^a)$ by suitably
modifying their lengths [by shifting the two waveforms in
  time, and by  truncating or extending $\hh(X^i+\Delta X^i,\Xi^a)$ at
  its starting point].}

To obtain the projected metric $g_{ij}^\mathrm{proj}$, we rewrite the
mismatch $\delta(\lambda^A,\lambda^A+\Delta \lambda^A)$ by separating
intrinsic and extrinsic parameters,
\begin{multline}
\delta(X^i,\Xi^\alpha;X^i+\Delta X^i,\Xi^\alpha+\Delta \Xi^\alpha) \\
= \left(
\begin{array}{cc}
\Delta X^i & \Delta \Xi^\alpha 
\end{array}\right)
\left(
\begin{array}{cc}
G_{ij} & C_{i \beta} \\
C_{\alpha j} &  \gamma_{\alpha\beta}
\end{array}\right)
\left(
\begin{array}{c}
\Delta X^j \\ \Delta \Xi^\beta
\end{array}\right);
\label{metric}
\end{multline}
here we have split the full metric $g_{BC}$ into four sections
corresponding to intrinsic--intrinsic ($G_{ij}$), extrinsic--extrinsic
($\gamma_{\alpha\beta}$), and mixed ($C_{\alpha j} = C_{j \alpha}$)
components. Maximizing the overlap over the extrinsic parameters is
then equivalent to minimizing Eq.\ \eqref{metric} over the $\Delta
\Xi^\alpha$ for a given $\Delta X^i$, which is achieved when
\begin{equation}
\gamma_{\alpha \beta} \Delta\Xi^{\beta} +  C_{\alpha j} \Delta X^j =0\,,
\label{extfromint}
\end{equation}
while the resulting mismatch is
\begin{multline}
\label{projmetric}
\min_{\Delta \Xi^\alpha} \delta(X^i,\Xi^\alpha;X^i+\Delta X^i,\Xi^\alpha+\Delta \Xi^\alpha) = \\
\begin{aligned}
& = 
\left[G_{ij}-C_{i\alpha}(\gamma^{-1})^{\alpha\beta}C_{\beta j}\right]
\Delta X^i \Delta X^j \\
& \equiv {g}_{ij}^{\rm proj} \Delta X^i \Delta X^j.
\end{aligned}
\end{multline}
Here $(\gamma^{-1})^{\alpha\beta}$ is the matrix inverse of
$\gamma_{\alpha\beta}$. For each point $(X^i,\Xi^\alpha)$ in the
\emph{full} parameter space, the \emph{projected metric}
$g^\mathrm{proj}_{ij}$ describes a set of concentric ellipsoids of
constant $\rho_{\Xi^\alpha}$ in the intrinsic-parameter
subspace. We emphasize that the projected metric has tensor indices
corresponding to the intrinsic parameters, but it is a function of
both the intrinsic and the extrinsic parameters, and so are the
constant-$\rho_{\Xi^\alpha}$ ellipsoids.

Therefore, to build a template bank that covers all the signals (for
all $X^i$ and $\Xi^\alpha$) with a guaranteed MM, we must use the
projected metric at each $X^i$ to construct the constant-mismatch
ellipsoids for all possible $\Xi^\alpha$, and then take the
intersection of these ellipsoids to determine the size of the unit
template-bank cell. This is a {\it minimax} prescription \cite{DIS}, because we
are maximizing the overlap over the extrinsic parameters of the
templates, and then setting the template-bank spacing according to the
least favorable extrinsic parameters of the signal.  In general, the
intersection of constant-mismatch ellipsoids is not an ellipsoid, even
in the limit $\delta \rightarrow 0$, so it is impossible to find a
single intrinsic-parameter metric that can be used to enforce the
minimax prescription.  There is an exception: the projected metric is
not a function of $t_0$ or $\Phi_0$,\footnote{The overlap $\langle
  \hh(X^i,\Xi^\alpha), \hh({X'}^i,{\Xi'}^\alpha) \rangle$ depends only
  on $t_0 - t'_0$ and $\Phi_0 - \Phi'_0$: we have $\hh(f) \sim \exp
  \{2\pi i f t_0 +i \Phi_0\}$ and $\hh'(f) \sim \exp \{2\pi i f t'_0
  +i\Phi'_0\}$, so $\hh'(f) \tilde{h}^*(f) \sim \exp \{2\pi i
  f(t'_0-t_0) + i (\Phi'_0-\Phi_0)\}$.} so it can be used directly to
lay down banks of nonspinning-binary templates \cite{BSD,O}, for which
$t_0$ and $\Phi_0$ are the only extrinsic parameters.

Returning to the generic case, we can still use the projected metric
to guide the placement of a template bank if we relax the minimax
prescription and request that the minimum match be guaranteed \emph{on
the average} for a distribution of signal extrinsic parameters.  It
turns out that this \emph{average-mismatch} prescription is closely
related to the expected detection rates. Let us see how.  The
matched-filtering detection rate for a signal $s \equiv
\mathrm{SA} \times
\hat{h}(X^i,\Xi^\alpha)$, where $\mathrm{SA} = \langle s, s
\rangle^{1/2}$ is the \emph{signal amplitude} at a fiducial luminosity
distance, is proportional to $\mathrm{SA}^3 \rho^3_{\Xi^\alpha}\![\hat{s},
\hat{h}_\mathrm{near}]$, where $\hat{h}_\mathrm{near} \equiv
\hat{h}(X^i+\Delta X^i,{\Xi'}^\alpha)$ is the closest template in the
bank, and where we assume that sources are uniformly distributed
throughout the volume accessible to the detector (see, for instance,
BCV1).  The minimax prescription is given by
\begin{equation}
\rho_{\Xi^\alpha}\![\hat{s},
\hat{h}_\mathrm{near}] \simeq
1 - g^\mathrm{proj}_{ij}(X^i, \Xi^\alpha) \Delta X^i \Delta X^j
\geq \mathrm{MM}
\end{equation}
for all $\Xi^\alpha$, which ensures that the detection rate is reduced
at most by a factor $\mathrm{MM}^3$ for every combination of
signal extrinsic and intrinsic parameters.

Averaging over a uniform distribution of signal extrinsic
  parameters,\footnote{All the expressions to follow can be adapted to
  the case of \emph{a priori} known probability distribution for the
  extrinsic parameters. However, in our case it seems quite right to
  assume that the orientation angles $\Theta$ and $\varphi$ are
  distributed uniformly over a sphere, and that $\alpha$ is
  distributed uniformly in the interval $[0,2\pi]$.} we get a
  detection rate proportional to
\begin{equation}
\begin{split}
\int \! d \Xi^\alpha \mathrm{SA}^3 \rho^3_{\Xi^\alpha}
&\simeq \int \! d \Xi^\alpha \mathrm{SA}^3 \left( 1 - g^\mathrm{proj}_{ij}
\Delta X^i \Delta X^j \right)^3 \\
&\simeq \overline{\mathrm{SA}^3} - 3 \left[ \int \! d \Xi^\alpha \mathrm{SA}^3
g^\mathrm{proj}_{ij} \right] \Delta X^i \Delta X^j \\
& \simeq \overline{\mathrm{SA}^3} \left( 1 -
\overline{g^\mathrm{proj}_{ij}} \Delta X^i \Delta X^j \right)^3, 
\end{split}
\label{eq:projmet}
\end{equation}
where $\overline{\mathrm{SA}^3} = \int d\Xi^\alpha \mathrm{SA}^3$, and
where the \emph{average metric} $\overline{g^\mathrm{proj}_{ij}}$, now
a function only of $X^i$, is defined as
\begin{equation}
\label{gavg}
\overline{g^\mathrm{proj}_{ij}} = \int d \Xi^\alpha \mathrm{SA}^3
g^\mathrm{proj}_{ij} \Big/ \, \overline{\mathrm{SA}^3}.
\end{equation}
[To derive Eq.\ \eqref{eq:projmet} we assume that $1 -
\rho_{\Xi^\alpha}\![\hat{s},\hh_\mathrm{near}] \ll 1$ for all $\Xi^\alpha$.]
We can now state the new average-mismatch prescription as
\begin{equation}
1 - \overline{g^\mathrm{proj}_{ij}}(X^i) \Delta X^i \Delta X^j
\geq \overline{\mathrm{MM}},
\end{equation}
which ensures that the detection rate, \emph{averaged over the
  extrinsic parameters of the signal}, is reduced at most by the
  factor $\overline{\mathrm{MM}}^3$. We shall call
  $\overline{\mathrm{MM}}$ the
\emph{average minimum match}.

\subsection{Null parameter directions and reduced metric}
\label{sec:reduction}

As discussed by Sathyaprakash and Schutz \cite{SS} and by Cutler
\cite{cpc}, an extreme example of boundary effects occurs when one of
the eigenvalues of $g_\mathrm{BC}$ at $\lambda^A$ (say,
$\Lambda_{(1)}$) becomes so small that it is possible to move away in
parameter space along the corresponding eigendirection (say,
$e_{(1)}^A$) and \emph{reach the boundary of the allowed parameter
region while keeping the mismatch $\delta(\lambda^A,\lambda^A + \tau
\, e_{(1)}^A)$ well below the required value} $\delta_\mathrm{MM} = 1
- \mathrm{MM}$. In other words, the ellipsoid of constant mismatch
$\delta_\mathrm{MM}$ extends far beyond the allowed parameter region
in the quasinull-eigenvalue direction. In such a situation, Eq.\
\eqref{calN} will underestimate the total number of templates, because
the denominator should now express the volume of the intersection of
each lattice cell with the allowed parameter
region.\footnote{Pictorially, the error that we make with Eq.\
\eqref{calN} is to let the template bank be thinner than a single
template in the direction $e_{(1)}^A$.} A simple-minded fix to Eq.\
\eqref{calN} is the following: write $\det g_{BC} = \prod_{(k)}
\Lambda_{(k)}$, where the $\Lambda_{(k)}$ are the $n$ eigenvalues of
$g_{BC}$; identify all the small eigenvalues, where
\emph{small} can be defined by $\Lambda_{(i)} \ll
(1-\mathrm{MM})/l^2_{(i)}$, with $l_{(i)}$ the coordinate diameter of
the allowed parameter range along the eigenvector $e_{(i)}^A$; replace
the small eigenvalues by the corresponding value of the expression
$(1-\mathrm{MM})/l^2_{(i)}$; use this modified determinant in Eq.\
\eqref{calN}.

Physically, the presence of $k$ small eigenvectors suggests that the
variety of waveform shapes spanned by an $n$-dimensional template
family can be approximated with very high overlap by an $(n-k)$-dim.\ 
\emph{reduced} family. A lower-dimensional template bank is certainly
desirable for practical purposes, but it is necessary to exercise
caution: because the metric $g_\mathrm{BC}$ is not homogeneous, the
quasinull eigendirections rotate as we move in parameter
space,\footnote{In fact, in the context of our templates this rotation
is such that Eq.\ \eqref{eq:basicmetric} ceases to be true in the
quasinull eigendirections for $\delta \gtrsim 0.01$. As soon as we
move away from the point $\lambda^A$ where the metric is computed, any
rotation of the eigenvectors means that the original quasinull
direction is no longer the path along which the mismatch grows most
slowly. If the larger eigenvalues are several orders of magnitude
larger than the smaller ones, as is true in our case, a tiny rotation
is enough to mask the contribution from the smallest eigenvalue.} so
we need to show explicitly that any signal in the $n$-dim.\ family can
be reached from a given $(n-k)$-dim.\ submanifold along a quasinull
trajectory.  For this to happen, the small eigenvalues must exist
throughout the entire $n$-dim.\ parameter space, and the flow of the
quasinull eigenvectors must map the submanifold into the entire
space. To see that under these conditions the mismatch between the
points on the submanifold and the points outside is indeed small,
consider the following argument, due to Curt Cutler \cite{cpc}.
The triangle inequality for the inner-product
distance guarantees that
\begin{equation}
\delta^{1/2}[\lambda^A(0), \lambda^A(1)] \leq
\int_{0}^{1} \sqrt{ g_{BC}
\frac{d\lambda^B}{d\nu} \frac{d\lambda^C}{d\nu}} d\nu
\end{equation}
along \emph{any} path $\lambda^A(\nu)$; for a path that follows the
flow of the quasinull eigenvector $e_{(i)}^A$ (a \emph{reduction
curve}), the total mismatch is then bounded by the average of
$\Lambda_{(i)}$ along the curve, times an integrated squared parameter
length of order $l_{(i)}$.\footnote{At least if the geometry of the
reduction curve is not very convoluted.}
\begin{figure}
\vspace{0.5cm}
\begin{center}
\includegraphics{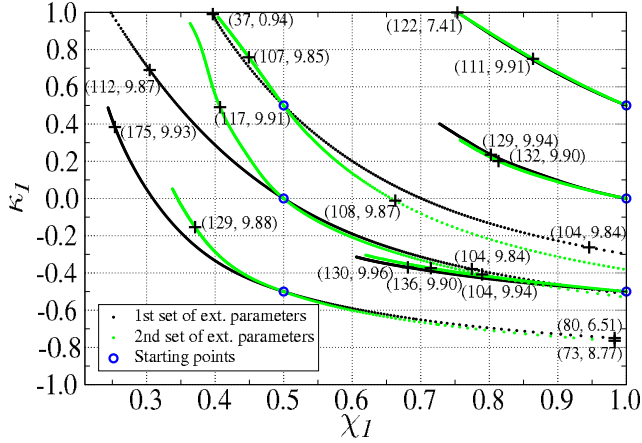}
\caption{
\label{redcurve}Plot of $(\chi_1,\kappa_1)$ reduction curves in the $(\chi_1,\kappa_1)$
plane. We show curves for two sets of starting extrinsic parameters,
corresponding to detector directions perpendicular (dark dots) and
parallel (light dots) to the initial orbital plane. The curves start
at the points marked with circles, and proceed in steps of $10^{-6}$
for the nominal mismatch (i.\ e., the mismatch computed using the
projected metric). For starting points at $\chi_1=0.5$, we follow the
quasi\-null eigenvector for both positive and negative increments.  The
curves end at the $(\chi_1,\kappa_1)$ boundary, or (roughly) where the
true mismatch (i.\ e., the exact mismatch between the local and
the starting template) becomes greater than $0.01$.  The ending points
are marked with crosses, and they are annotated with the number of
steps taken since the starting point, and with the true mismatch in
units of $10^{-3}$.}
\end{center}
\end{figure}
\begin{figure}
\vspace{0.5cm}
\begin{center}
\includegraphics{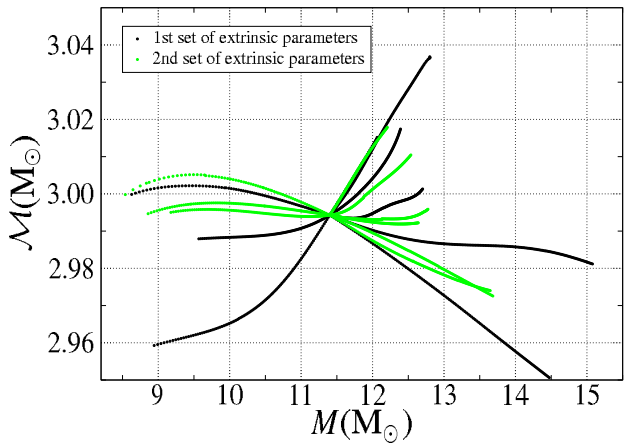}
\caption{\label{redcurvemass}Plot of $(\chi_1,\kappa_1)$ reduction
curves in the $(M,\mathcal{M})$ plane. The curves are the same as
shown in Fig.\ \ref{redcurve}, but we omit all markings.}
\end{center}
\end{figure}

For the $\mathrm{ST}_N$ template bank and for the two-stage search
scheme of Sec.\ \ref{sec:testing}, we find that the projected metric
$g^\mathrm{proj}_{ij}$ admits a small eigenvalue for all values of the
intrinsic and extrinsic parameters. Figures \ref{redcurve} and
\ref{redcurvemass} show several examples of reduction curves that
follow the quasinull eigendirections [the subtleties related to
projected-metric reduction curves are discussed in App.\
\ref{app:reduction}].  The curves shown\footnote{The curves of Figs.\
\ref{redcurve} and \ref{redcurvemass}
  are in fact obtained by following the quasinull eigenvectors of the
  \emph{fully projected} $(\chi_1,\kappa_1)$ metric, which is a 2-D
  metric on $\chi_1$ and $\kappa_1$ obtained by projecting
  $g^\mathrm{proj}_{ij}$ again over $M$ and $\eta$ [using
  Eq.~\eqref{projmetric}], {\it as if they were extrinsic parameters.}
  The two projection steps are equivalent to projecting the 9-D full
  metric into the 2-D $(\chi_1,\kappa_1)$ plane in a single step.
  This procedure estimates correctly the \emph{reduction mismatch}
  introduced by adopting the reduced template family created by first
  collapsing the 2-D $(\chi_1,\kappa_1)$ plane into several 1-D
  curves, and then including all the values of $(M,\eta)$ for each
  point on the curves. We choose to collapse the $(\chi_1,\kappa_1)$
  plane for empirical reasons: the $\chi_1$ and $\kappa_1$ parameter
  bounds are simple, $(\chi_1,\kappa_1) \in [0,1]\times[-1,1]$, and
  the reduction curves have large parameter lengths.} begin at the
points marked with circles, where $(m_1+m_2)=(10+1.4)M_\odot$ and
\begin{equation}
(\chi_1, \kappa_1) = 
\left\{
\begin{array}{r}
0.5 \\
1.0 \\
\end{array}
\right\} \times
\left\{
\begin{array}{c}
-0.5 \\ 0.0 \\ 0.5
\end{array}
\right\};
\end{equation}
the curves then proceed in steps of $10^{-6}$ for the \emph{nominal}
mismatch (i.\ e., the mismatch computed using the local projected metric)
until they reach the $(\chi_1,\kappa_1)$ boundary, or (roughly) until
the \emph{true} mismatch (i.\ e., the exact mismatch between the local
and the starting template) is greater than $0.01$. We show curves for
two sets of starting extrinsic parameters, corresponding to detector
directions perpendicular (dark dots) and parallel (light dots) to the
initial orbital plane. Figure \ref{redcurve} shows the projection of
the reduction curves in the $(\chi_1,\kappa_1)$ plane; the ending
points are marked with crosses, and they are annotated with the number
of steps taken since the starting point, and with the true mismatch in
units of $10^{-3}$. Comparing the two numbers at each cross, we see that
the triangle inequality is always respected: the true mismatch
$\delta_N$ is always less than the accumulated nominal mismatch
$10^{-6} N^2$ (where $N$ is the number of steps); in fact, we see that
the latter is a good approximation for the former. Figure
\ref{redcurvemass} shows the projection of the same reduction curves
in the $(M,\mathcal{M})$ plane. The chirp mass $\mathcal{M}\equiv
M\eta^{3/5}$ varies by less than 2\% along the curves: this is
natural, since $\mathcal{M}$ dominates the evolution of the GW
phase [see Eq.~\eqref{omegadot}].

Figure \ref{redcurve} suggests that we can reduce the dimensionality
of our template bank by collapsing the $(\chi_1,\kappa_1)$ plane into
$\sim$ three curves, while retaining the full $(M,\eta)$ plane.
Templates laid down on these 3-D submanifolds with a required minimum
match MM will then cover every signal in the full 4-D family with
mismatch no larger than $(1-\mathrm{MM}) + \delta_\mathrm{red}$, where
$\delta_\mathrm{red} \simeq 0.01$ is the \emph{reduction mismatch}
introduced by the reduction procedure.
Further investigations will be needed to find the optimal choice of
reduction curves in the $(\chi_1,\kappa_1)$ plane, and to investigate
the reduction curves of the average metric $\overline{g_{ij}^{\rm
proj}}$.

\subsection{Template counting}
\label{sec:counting}

While three or more reduction curves will probably be necessary to
limit $\delta_\mathrm{red} \simeq 0.01$, for the sake of
definiteness we select a 3-D reduced template space corresponding to
$(m_1,m_2) \in [1,3] \times [7,12]$, $\kappa_1=0$, and $\chi_1 \in
(0,1].$\footnote{In fact, a second small eigenvector appears as we
  get close to $\chi_1 = 0$; this is because spin effects vanish in
  that limit, so a 2-D family of nonspinning waveforms should be
  sufficient to fit all signals with small $\chi_1$.} We compute the
total number of templates in this 3-D template bank according to
\begin{equation}
{\cal N}_\mathrm{templates}=
\frac{ {\displaystyle\int} \! \sqrt{\left|\det \overline{g^\mathrm{proj}_{i'j'}}\right|} dM \, d\eta \, d\chi}{
\left[2 \sqrt{(1-{\rm MM})/3}\right]^3},
\end{equation}
where the primed indices $i'$, $j'$ run through $M$, $\eta$, and
$\chi$, and we set $X^4 \equiv \kappa_1 = 0$; furthermore,
$\overline{g^{\rm proj}_{ij}}$ denotes the metric averaged over the
extrinsic parameters $\Theta$, $\varphi$, and $\alpha$, as given by
Eq.~\eqref{gavg}.  The integral is carried out by evaluating the
projected metric at the parameter sets
\begin{equation}
(m_1, m_2, \chi_1) = 
\left\{
\begin{array}{r}
7 M_\odot \\
12 M_\odot
\end{array}
\right\} \times
\left\{
\begin{array}{r}
1 M_\odot \\
2 M_\odot \\
3 M_\odot
\end{array}
\right\} \times
\left\{
\begin{array}{c}
0.1 \\ 0.3 \\ 0.5 \\ 0.7 \\ 1.0
\end{array}
\right\};
\label{eq:parsets}
\end{equation}
at each of the points the metric is averaged on 100 pseudorandom sets
of extrinsic parameters. The integration then proceeds by
interpolating across the parameter sets \eqref{eq:parsets}. The final
result is $\mathcal{N}_\mathrm{templates} \simeq$ 76,000 for
$\mathrm{MM} = 0.98$ (not including the reduction mismatch). Given the
uncertainties implicit in the numerical computation of the metric, in
the interpolation, in the choice of the reduction curves, and in the
actual placement of the templates in the bank, this number should be
understood as an order-of-magnitude estimate.  Most of the templates,
by a factor of about ten to one, come from the parameter region near
$m_2 = 1$ (that is, from the small-$\eta$ region).

\section{Summary}
\label{sec:conclusion}

Buonanno, Chen, and Vallisneri recently proposed [BCV2] a family of
physical templates that can be used to detect the GWs emitted by
single-spin precessing binaries. The attribute \emph{physical} refers
to the fact that the templates are exact within the approximations
used to write the PN equations that rule the adiabatic evolution of
the binary.  In this paper, after reviewing the definition of this
template family (here denoted as $\mathrm{ST}_N$), we discuss the
range of binary masses for which the templates can be considered
accurate, and examine the effects of higher-order PN corrections,
including quadrupole-monopole interactions. We then describe an
optimized two-stage detection scheme that employs the $\mathrm{ST}_N$
family, and investigate its false-alarm statistics. Finally, we
estimate the number of templates needed in a GW search with LIGO-I.
Our results can be summarized as follows.

We determine the range of binary masses where the $\mathrm{ST}_N$
templates can be considered accurate by imposing two conditions:
first, for the orbital separations that correspond to GWs in the
frequency band of good interferometer sensitivity, the dynamics of the
binary must be described faithfully by an adiabatic sequence of
quasi-spherical orbits; second, the nonspinning body must be light
enough that its spin will be negligible for purely dimensional reasons.
The selected mass range  is $(m_1,m_2) \simeq [7,12]M_\odot \times [1,3]M_\odot$.

To evaluate the effect of higher-order PN corrections for binaries in
this mass range, we compute the overlaps between templates computed at
successive PN orders. When computed between templates with the same
parameters, such overlaps can be rather low; however, they become very
high when maximized over the parameters (both intrinsic and extrinsic)
of the lower-order PN template [see Table \ref{CauchyST123}]. This
means that the $\mathrm{ST}_2$ template family 
should be considered acceptable for the purpose of GW detection;
but this means also that the estimation of certain combinations of
binary parameters can be affected by large systematic
errors~\cite{pbcv3}.  [When precessing-binary gravitational waveforms
computed within PN-resummed and nonadiabatic approaches \cite{DIS,EOB}
become available, it will be interesting to compare them with the
PN-expanded, adiabatic $\mathrm{ST}_N$ templates, to see if the
maximized overlaps remain high.  We do expect this to be the case,
because the spin and directional parameters of the $\mathrm{ST}_N$
templates provide much leeway to compensate for nontrivial variations
in the PN phasing.] Again by considering maximized overlaps, we
establish that quadrupole--monopole effects \cite{G,QM} can be safely
neglected for the range of masses investigated [Table~\ref{tab:QM}].

We describe a two-stage GW detection scheme that employs a discrete
bank of $\mathrm{ST}_2$ templates laid down along the intrinsic
parameters $(M, \eta, \chi_1, \kappa_1)$ [although the
$(\chi_1,\kappa_1)$ may be collapsed to one or few 1-D curves, in
light of the discussion of dimensional reduction of Sec.\ 
\ref{sec:metric}].  The detection statistic $\rhocon(M, \eta, \chi_1,
\kappa_1)$ is the overlap between the template and the detector
output, maximized over template extrinsic parameters: $(t_0, \Phi_0,
P_I) \equiv (t_0, \Phi_0, \theta, \phi, \psi, \Theta, \varphi)$. This
maximization is performed semialgebraically, in two stages. First, for
all possible times of arrival $t_0$, we maximize the overlap over
$\Phi_0$ and over $P_I$ without accounting for the constraints that
express the functional dependence of the $P_I$ on $(\theta, \phi,
\psi, \Theta, \varphi)$: this step yields the approximated
(unconstrained) maximum $\rhounc$, which can be computed very rapidly,
and which sets an upper bound for $\rhocon$. Second, only for the times of
arrival $t_0$ at which $\rhounc$ passes the detection threshold, we
compute the fully constrained maximum $\rhocon$, which is more
expensive to compute. [Note that this scheme differs from traditional
hierarchical schemes because we use the {\it same} threshold in the first
and second stages.]  We find that $\rhounc$ is a good approximation to
$\rhocon$, so the number of first-stage triggers passed to the second
stage is small.

For a total false-alarm probability of $10^{-3}$/year, and for a
conservative estimate for the number of independent statistical tests,
the detection threshold is around 10.  For this value, between 5 and
15 first-stage triggers are passed to the second stage for each
eventual detection. For the same threshold, the single-test
false-alarm probability is lower for $\mathrm{ST}_2$ templates than
for the $(\psi_0 \psi_{3/2} \mathcal{B})_6$ DTF of BCV2 [the total
false-alarm probability depends on the number of independent
statistical tests, which is not available at this time for the
$(\psi_0 \psi_{3/2} \mathcal{B})_6$ DTF].

The procedure of maximization over the extrinsic parameters outlined
in this paper can also be adapted for the task of detecting GWs from
extreme--mass-ratio inspirals (i.\ e., the inspiral of solar-mass
compact objects into the supermassive BHs at the center of galaxies
\cite{extreme}) and inspirals of two supermassive black holes
with LISA \cite{LISA}. This is possible under the simplifying
assumptions of coherent matched filtering over times short enough that
the LISA antenna patterns can be considered constant, and of GW
emission described by the quadrupole formula. Furthermore, the
formalism of projected and reduced mismatch metrics developed in Sec.\
\ref{sec:metric} can treat GW sources, such as extreme--mass-ratio
inspirals, where many physical parameters are present, but only few of
their combinations have significant effects on the emitted waveforms
\cite{SS,cpc}. In fact, this formalism is closely related to the
procedures and approximations used in the ongoing effort (motivated by
mission-design considerations) to count the templates needed to detect
extreme--mass-ratio inspirals with LISA
\cite{LISAcounting}.

It should be possible to generalize the formalism beyond quadrupole GW
emission, at least to some extent.  When higher-multipole
contributions are included, the detector response becomes much more
complicated than Eq.~(\ref{generich}) (see, e.\ g., Eqs.
(3.22b)--(3.22h) of Ref.~\cite{K}). In particular, the response cannot
be factorized into a factor that depends only on the dynamical
evolution of the binary, and a factor that depends only on the
position and orientation of the detector; it is is instead a sum over
a number of such terms, each containing different harmonics of the
orbital and modulation frequencies.  Despite these complications, it
should still be possible to maximize the overlap over the extrinsic
parameters, using a relatively small number of signal--template and
template--template inner products.  The constrained-maximization
procedure would however be very complicated, and although the (fully
algebraic) unconstrained maximum would still be easy to compute, the
dimensionality of the unconstrained template space would now be so
large that it may increase the false alarm probability too
dramatically to make the two-stage scheme useful.

The last result of this paper is an estimate of the number of
$\mathrm{ST}_2$ templates needed for a GW search in the mass range
$[7,12]M_\odot \times [1,3]M_\odot$. To obtain this estimate, we first
compute the full mismatch metric, which describes the mismatch for
small displacements in the intrinsic and extrinsic parameters; we then
obtain the projected metric, which reproduces the effect of maximizing
the overlap over the extrinsic parameters. At this point we observe
that the projected metric has an eigenvector corresponding to a very
small eigenvalue; this indicates that we can choose one of the four
intrinsic parameters to be a function of the other three, so the
dimensionality of the $\mathrm{ST}_2$ template bank can be reduced to
three. For simplicity, we perform this reduction by setting $\kappa_1
= 0$. We then compute the reduced mismatch metric, and obtain a rough
estimate of $\sim 76,000$ as the number of templates required for an
average MM of 0.98, or 0.97 including an estimated reduction mismatch
of 0.01.

\acknowledgments We wish to thank Leor Barack, Teviet Creighton, Curt
Cutler and Jonathan Gair for useful discussions.  A.\ B.\ thanks the
LIGO Caltech Laboratory under NSF cooperative agreement PHY92-10038
for support and the Theoretical Astrophysics group for hospitality
during her visit at Caltech.  The research of Y.~C.\ and Y.~P.\ was
supported by NSF grant PHY-0099568 and NASA grant NAG5-12834. Y.~C.\ 
is also supported by the David and Barbara Groce Fund at the San Diego
Foundation.  M.\ V.\ performed this research at the Jet Propulsion
Laboratory, California Institute of Technology, under contract with
the National Aeronautics and Space Administration.

\appendix

\section{Algebraic maximization of the overlap over the $P_I$}
\label{app:maximize}

In this section, we explore the algebraic maximization of $\rho_{\Phi_0}$
[see Eq.~\eqref{rhoapp}], given by
\beq
\label{rhoapp_app}
\rho_{\Phi_0} 
= \sqrt{\frac{A^{IJ}P_I P_J}{
    B^{IJ}P_I P_J}}\,,
\eeq
over the $P_I$. We recall that the five $P_I$ are combinations of
trigonometric functions of three angles, and therefore must satisfy
two constraints: luckily, both of these can be formulated
algebraically. In light of the discussion of
Sec.~\ref{subsec:maximize}, the overall normalization of the $P_I$
does not affect the value of the overlap~\eqref{rhoapp}. As a
consequence, we can rescale the $P_I$ and replace the first constraint
by
\beq
\label{eq:normalize}
B^{IJ}P_{I}P_{J}  = 1\,,
\eeq
which enforces the normalization of the templates. This constraint is
chosen only for convenience: the maximum, subject to this
constraint, is exactly the same as the unconstrained maximum found by
searching over the entire five-dimensional space.
Let us work out its value, which will be useful later.
Introducing the first Lagrangian multiplier $\lambda$, we
impose
\beq
\label{PIunconstrained}
\frac{\partial}{\partial P_I} [ A^{IJ} P_I P_J - \lambda (
B^{IJ}P_I P_J - 1)] = (A^{IJ}-\lambda B^{IJ})P_J = 0 \,,
\eeq
which has solutions only for $\lambda$ corresponding to the
eigenvalues of $\mathbf{A} \mathbf{B}^{-1}$. For those solutions,
we multiply Eq.\ \eqref{PIunconstrained} by $P_I$ to obtain
\begin{equation}
\lambda = A^{IJ}P_{I}P_{J};
\label{forlambda}
\end{equation}
using Eqs.\ \eqref{rhoapp_app} and \eqref{eq:normalize}, we then see that
$\lambda$ is the square of the  
overlap, so it should be chosen as the largest eigenvalue of $\mathbf{A}\,
\mathbf{B}^{-1}$. We then write the \emph{unconstrained maximum} as
\beq
\label{unc:app}
\rho_{\Xi^\alpha}'=\max_{t_0}\sqrt{\max {\rm eigv}
\mathbf{A}\,\mathbf{B}^{-1}}\,. 
\eeq
By construction, $\rhounc$ will always be larger than or equal to the
 constrained maximum, $\rhocon$.

The second constraint comes from Eq.~(\ref{detP}). Writing out the STF
components, we get
\begin{widetext}
\begin{equation}
\label{detPIJK}
\det P_{ij} \equiv
\det \frac{1}{\sqrt{2}}
\begin{pmatrix}
P_1+P_5/\sqrt{3} &  P_2              &    P_3 \\
P_2              & -P_1+P_5/\sqrt{3} &    P_4 \\
P_3              &  P_4              & -2 P_5 / \sqrt{3}
\end{pmatrix}
\equiv D^{IJK} P_I P_J P_K = 0.
\end{equation}
\end{widetext}
[The tensor $D^{IJK}$ can be chosen to be symmetric since
$D^{IJK}P_{I}P_{J}P_{K} = D^{(IJK)}P_{I}P_{J}P_{K}$.]
The constrained maximum of $\rho_{\Phi_0}$ over the $P_I$, subject to
the two constraints, can be obtained as the maximum of the function 
\begin{equation}
\label{eq:Lag}
A^{IJ}P_{I}P_{J} - \lambda (B^{IJ}P_{I}P_{J}-1) - \mu (D^{IJK}P_{I}P_{J}P_{K})
\end{equation}
over $P_I$ and over the two Lagrange multipliers $\lambda$ and $\mu$.
After taking partial derivatives, we get a system of seven equations,
\begin{eqnarray}
\label{Peqns11}
A^{IJ}P_{J}-\lambda B^{IJ}P_{J}-\frac{3}{2}\mu D^{IJK}P_{J}P_{K}&=&0\,,\\
B^{IJ}P_{I}P_{J}-1&=&0\,,\label{Peqns12}\\
D^{IJK}P_{I}P_{J}P_{K}&=&0\,,\label{Peqns13}
\end{eqnarray}
where the last two equations come from the
constraints~\eqref{eq:normalize} and \eqref{detPIJK}.  Multiplying the
first equation by $P_{I}$ and using the two constraints, we obtain
Eq.~\eqref{forlambda} again.  So the first Lagrange multiplier
$\lambda$ is still the square of the overlap.
The second Lagrange multiplier $\mu$ is zero when the signal $s$
belongs to $\mathrm{ST}_N$ template family,
and has exactly the same intrinsic parameters as the template. In this
case, the extrinsic parameters of the signal correspond to a vector
$P_I$ that satisfies Eq.\ \eqref{Peqns11} with $\mu=0$ (the multiplier
$\lambda$ is still needed to enforce normalization of the
template). When the intrinsic parameters are not exactly equal, but
close, $\mu$ becomes finite, but small. 
Equations \eqref{Peqns11}--\eqref{Peqns13} can then be solved
iteratively by expanding $P_{I}$ in terms of $\mu$,
\begin{equation}
\label{eq:expansion}
P_{I}=\sum_{n=0}^{\infty} P_{I}^{(n)} \mu^{n}.
\end{equation}
Inserting this expansion into Eqs.\ \eqref{Peqns11} and
\eqref{Peqns13}, we get the zeroth-order equation
\begin{equation}
\label{zeroth1}
A^{IJ}P^{(0)}_{J}-(A^{LM}P^{(0)}_{L}P^{(0)}_{M})B^{IJ}P^{(0)}_{J}=0.
\end{equation}
where we have already used the zeroth-order version of Eq.\
\eqref{forlambda} to eliminate $\lambda$.

Multiplying by ${(B^{-1})^{KI}}$, we see that the zeroth-order
solution $P^{(0)}_{J}$ must lie along an eigenvector of
${(B^{-1})^{KI}} A^{IJ}$, and that the corresponding eigenvalue must
be equal to $A^{LM} P^{(0)}_L P^{(0)}_M$, and therefore also to the
square of the zeroth-order extremized overlap. To get the maximum
overlap, we must therefore choose $P^{(0)}_{I}$ as the eigenvector
corresponding to the largest eigenvalue.  So the zeroth-order
constrained maximum is exactly the unconstrained maximum obtained
above [Eqs.~\eqref{PIunconstrained}--\eqref{unc:app}].

We can then proceed to $n$th-order equations: 
\begin{equation}
\begin{split}
[A^{IJ}-&2(A^{JM}P^{(0)}_M B^{IL}P^{(0)}_L)-
(A^{LM}P^{(0)}_{L}P^{(0)}_{M})B^{IJ}] P_J^{(n)} \\
= & \sum_{m_1=0}^{n-1}\sum_{m_2=0}^{n-1} A^{LM}P^{(m_1)}_L P^{(m_2)}_M
B^{IJ}P^{(n-m_1-m_2)}_J + \\
& \sum_{m=0}^{n-1}\frac{3}{2}D^{IJK}P^{(m)}_J P^{(n-m-1)}_K.
\end{split}
\end{equation}
At each order, we insert the
$n$th-order expansion of $P_I$ into Eq.\
\eqref{Peqns13}, and select the real solution closest to zero
as the $n$th-order approximation to $\mu$ (such a solution is
guaranteed to exist for all odd $n$). We then obtain the $n$th-order
approximation to $\lambda$ (and therefore to $\rhocon$) using Eq.\
\eqref{forlambda}. We proceed in this way, until $\lambda$ and $\mu$
converge to our satisfaction.

This iterative procedure succeeds when the intrinsic parameters of
signal and template are close; as their distance increases, the
procedure becomes more and more unstable, and eventually fails to
converge. The iteration fails often also when the overlap is optimized
against pure noise. For these reasons, a practical implementation of
the detection statistic $\rho_{\Xi^{\alpha}}$ must eventually rely
on the semialgebraic maximization procedure discussed in Sec.\
\ref{subsec:maximize}. Indeed, we have used the semialgebraic
procedure for all the tests discussed in Sec.\ \ref{sec:testing}.

\section{Dimensional reduction with a nonuniform projected metric}
\label{app:reduction}

In this appendix we extend the reasoning of Sec.\ \ref{sec:reduction}
to study dimensional reduction under the projected metric $g^{\rm
proj}_{ij}(\lambda^A)$, which lives in the intrinsic parameter space,
but is a function of all parameters. For each point
$\lambda^A=(X^i,\Xi^\alpha)$ in parameter space, we denote
$\Lambda_{(1)}(\lambda^A)$ the smallest eigenvalue of $g^{\rm
proj}_{ij}(\lambda^A)$, and $e_{(1)}^i(\lambda^A)$ the corresponding
eigenvector in the intrinsic parameter space. Suppose we have
\begin{equation}
\Lambda_{(1)}(\lambda^A) \ll \frac{1-\mathrm{MM}}{l_{(1)}^2},
\end{equation}
for all values of $\lambda^A$ in the allowed parameter region, where
$l_{(1)}$ is the coordinate diameter of the allowed parameter range
along the eigenvector $e_{(1)}^i$.

Now let us start from a generic
point $\lambda_0^A = (X^i_0,\Xi^\alpha_0)$ in parameter space and
follow the normal eigenvector $e_{(1)}^i$ for a tiny parameter length
$\epsilon$, reaching $\lambda^A_1 = (X^i_1,\Xi^\alpha_1)$, according to
\bea
\label{Xdiff}
X_1^i & = &  X_0^i + \epsilon\, e_{(1)}^i(\lambda^A_0), \\
\label{Xidiff}
\Xi_1^i &=& \Xi_0^i + \epsilon 
\left[
\gamma^{-1}(\lambda^A_0)\right]^{\alpha\beta}\left[C(\lambda^A_0)\right]
_{\beta j} e_{(1)}^j(\lambda^A_0);
\eea
this choice of $\Delta\Xi^\alpha$ makes $\Xi_1^\alpha$ the extrinsic
parameter that minimizes
$\delta(X_0^i,\Xi_0^\alpha;X_1^i,\Xi_1^\alpha)$.  Denoting the
inner-product distance as $\mathrm{dist}(\lambda_0^A,\lambda_1^A) \equiv
\sqrt{2 \delta(\lambda_0^A,\lambda_1^A)}$, we can write
\beq
{\rm dist}(\lambda^A_0,\lambda^A_1) = \epsilon \sqrt{2 \Lambda_{(1)}(\lambda^A_0)} + O(\epsilon^2);
\eeq
from $\lambda^A_1$, we follow the eigenvector $e_{(1)}^i(\lambda_1^A)$
for another parameter length $\epsilon$, and reach $\lambda^A_2$; then
from $\lambda^A_2$ we reach $\lambda^A_3$, and so on. Up to the $N$th
step, we have traveled a cumulative parameter distance $l=N\epsilon$
in the intrinsic parameter space, and an inner-product distance
\bea
{\rm dist}(\lambda^A_{0},\lambda^A_N)
&\le&
\sum_{n=1}^N {\rm dist}(\lambda^A_{n-1},\lambda^A_{n}) \nonumber \\
&=& \sum_{n=1}^N  \left[\epsilon \sqrt{2 \Lambda_{(1)}(\lambda^A_{n-1})} +
O(\epsilon^2)\right] \nonumber \\
&\le& l\sqrt{2 \max_{\lambda^A} \Lambda_{(1)}(\lambda^A)}+ O(N\epsilon^2),
\label{distdiff}
\eea
where in the first line we have used the triangle inequality for the
inner-product distance.  The term $O(N\epsilon^2)$ vanishes in the
limit $\epsilon
\rightarrow 0$, $N\rightarrow \infty$, keeping $l=N\epsilon$
finite (see Fig.\ \ref{fig:reduction}). So we can take the continuous
limit of Eqs.~(\ref{Xdiff}) and (\ref{Xidiff}) and arrive at two
differential equations that define the resulting trajectory:
\begin{figure}
\begin{center}
\includegraphics[width=0.4\textwidth]{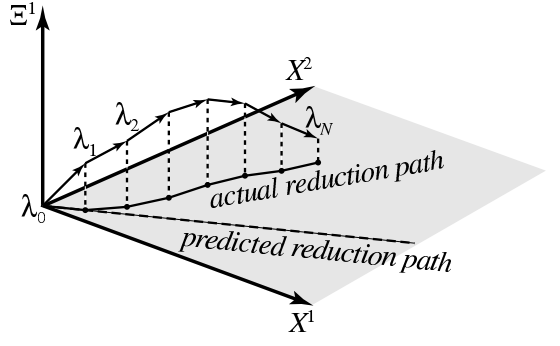}
\caption{Illustration of dimensional reduction. Here we show 
a signal space with one extrinsic parameter ($\Xi^1$) and two
intrinsic parameters ($X^{1,2}$), and we assume that the projected
metric has one small eigenvalue all through parameter space.  Starting
from a generic point $\lambda^A_0$, we follow the flow of the
quasinull eigenvector of $g_{ij}^{\rm proj}$ for an infinitesimal
parameter distance to reach $\lambda^A_1$; we then repeat this
process, each time adjusting the direction of the eigenvector
according to the metric (hence the difference between the reduction
path \emph{predicted} at $\lambda^A_0$ and the \emph{actual} reduction
path).  In the end we reach $\lambda^A_N$ after having accumulated a
parameter length $l$ in the \emph{intrinsic} parameter space. The
mismatch between $\lambda_0$ and $\lambda_N$ will be smaller than
$\delta_\mathrm{MM} = 1 - \mathrm{MM}$, if $l$ is not much larger than
$l_{(1)}$, the coordinate diameter of the intrinsic parameter space in
the approximate direction of the quasinull eigenvector.
\label{fig:reduction}}
\end{center}
\end{figure}
\begin{figure}
\begin{center}
\includegraphics[width=0.4\textwidth]{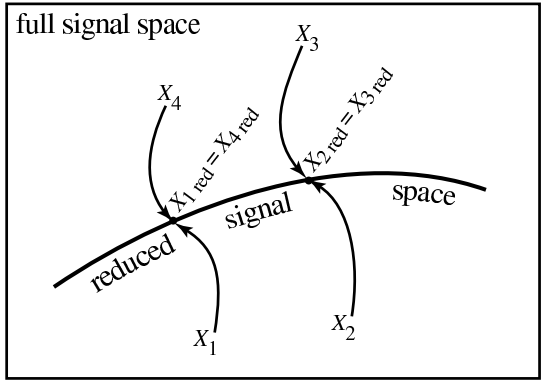}
\caption{Illustration of reduced signal space as a hypersurface
inside full signal space. Here we show only the directions along the
intrinsic parameters. Starting from the points $(X^i_1,\Xi^\alpha_1)$,
\ldots, $(X^i_4,\Xi^\alpha_4)$, we follow the trajectory (\ref{dlambda})
and reach the hypersurface at $(X^i_{1\,\rm red},\Xi^\alpha_{1\,\rm red})$,
\ldots, $(X^i_{4\,\rm red},\Xi^\alpha_{4\,\rm red})$.
For these particular initial points, $X_{1\,\rm red}$ happens to
coincide with $X_{4\,\rm red}$, and $X_{2\,\rm red}$ with $X_{3\,\rm
red}$. We can see that $\lambda^A_1$ and $\lambda^A_4$ (and indeed all
points that reduce to $X_{1\,\rm red}$, including the points along the
reduction curve) will be indistinguishable upon detection with the
reduced template bank. The same is true for $\lambda_2^A$, $\lambda_3^A$,
and for all the points that reduce to $X_{1\,\rm red}$.\label{fig:red2}}
\end{center}
\end{figure}
\beq
\label{dlambda}
\dot{X}^i(l) = e^i_{(1)},\quad
\dot{\Xi}^{\alpha}(l) = \left[\gamma^{-1}\right]^{\alpha\beta} 
C_{\beta j} e^{j}_{(1)},
\eeq
where $X^i$ and $\Xi^{\alpha}$ are parametrized by the cumulative
parameter length $l$, with
\beq
X^i(l=0)=X^i_0\,,\qquad \Xi^\alpha(l=0)=\Xi^\alpha_0\,.
\eeq
We can allow $l$ to be either positive or negative, in order to
describe the two trajectories that initially propagate along $\pm
e^i_{(1)}(\lambda^A_0)$. Equation (\ref{distdiff}) then becomes
\bea
\label{ddist}
{\rm dist}\left[\lambda^A_0,\lambda^A(l)\right] &\le& \int_0^l dl'
\sqrt{2 \Lambda_{(1)}\left[\lambda^A(l')\right]} \nonumber \\
&\le& |l|\sqrt{2 \max_{\lambda^A} \Lambda_{(1)}(\lambda^A)}.
\eea
In terms of mismatch,
\bea
\min_{\Xi^\alpha} \delta \left[\lambda^A_0;X^i(l)\right]
&=&\frac{1}{2}\Big[\min_{\Xi^\alpha} {\rm dist}
\left[\lambda^A_0;X^i(l)\right] \Big]^2 \nonumber \\
&\le &\frac{1}{2}\Big[ {\rm dist}\left[\lambda^A_0;\lambda^A(l)\right] \Big]^2 \nonumber \\
 &\le& \frac{1}{2}\left[ \int_0^l dl'
\sqrt{2 \Lambda_{(1)}\left[\lambda^A(l')\right]}\right]^2 \nonumber
 \\
&\le& 
l^2 
\max_{\lambda^A} \Lambda_{(1)}(\lambda^A) \nonumber \\
&\ll& \left(\frac{l}{l_{(1)}}\right)^2 \delta_\mathrm{MM},
\label{deltareduce}
\eea
where the hybrid notation of the first line indicates the mismatch
along the solution of \eqref{dlambda}, and where of course
$\delta_\mathrm{MM} = 1 - \mathrm{MM}$.  Here, although we evolve
$X^i$ and $\Xi^\alpha$ simultaneously, it is the trajectory $X^i(l)$
in the \emph{intrinsic} parameter space that we are ultimately interested
in. In the context of dimensional reduction for the projected metric,
we shall call $X^i(l)$ the reduction curve.

If the reduction curves are reasonably straight, it should be easy to
find a (dimensionally reduced) hypersurface with the property that any
given point $(X^i_0,\Xi^{\alpha}_0)$ in the full parameter space
admits a reduction curve that reaches the hypersurface at a parameter
$l_*$ not much larger than the coordinate diameter of parameter space
(see Fig.\ \ref{fig:red2}). From Eq.\ (\ref{deltareduce}), we then
have $\min_{\Xi^\alpha} \delta[X^i_0,\Xi^{\alpha}_0;X^i(l_*)] <
\delta_\mathrm{MM}$. So any point in the full parameter space can be fit 
with a mismatch smaller than $\delta_\mathrm{MM}$ by a point on the
hypersurface.

\end{document}